%% file: paper.tex
\documentclass{JHEP3}
\usepackage{epsfig}
\usepackage{subfigure}
\usepackage{psfrag}
\usepackage{graphicx}

\input texpref.tex

\newcommand{\be}{\begin{equation}}
\newcommand{\ee}{\end{equation}}
\newcommand{\bea}{\begin{eqnarray}}
\newcommand{\eea}{\end{eqnarray}}
\newcommand{\bean}{\begin{eqnarray*}}
\newcommand{\eean}{\end{eqnarray*}}

\newcommand{\beqa}{\begin{eqnarray}}
\newcommand{\eeqa}{\end{eqnarray}}
\newcommand{\id}{\bf 1}

\def\tr{{\rm tr \,}}

\def\id {{\bf 1}}

\preprint{SISSA-62/2006/EP \\ PUPT-2212 \\SLAC-PUB-12130}

\title{Gauge/gravity duality and meta-stable \\ dynamical supersymmetry breaking}

\author{Riccardo Argurio$^1$, Matteo Bertolini$^2$,
Sebasti\'an Franco$^3$ and Shamit Kachru$^{4,5}$

\\
~\\

${}^1$Physique Th\'eorique et Math\'ematique and International Solvay Institutes \\
Universit\'e Libre de Bruxelles, C.P. 231, 1050 Bruxelles, Belgium \\ \vspace{0.3cm}
${}^2$SISSA/ISAS and INFN - Sezione di Trieste \\
Via Beirut 2; I 34014 Trieste, Italy \\ \vspace{0.3cm}
${}^3$Joseph Henry Laboratories, Princeton University \\
Princeton, NJ 08544, USA \\ \vspace{0.3cm}
${}^4$Department of Physics and SLAC, Stanford University \\
Stanford, CA 94305 USA \\ \vspace{0.3cm}
${}^5$Kavli Institute for Theoretical Physics, University of California \\
Santa Barbara, CA 93106 USA
\\

\email{rargurio@ulb.ac.be, bertmat@sissa.it, sfranco@feynman.princeton.edu,
skachru@stanford.edu}\\
}

\abstract{
We engineer a class of quiver gauge theories with several interesting
features by studying D-branes at a simple Calabi-Yau
singularity. At weak 't Hooft coupling we argue using field theory
techniques that these theories 
admit both supersymmetric vacua and meta-stable
non-supersymmetric vacua, though the arguments 
indicating the existence of the supersymmetry breaking
states are not decisive. At strong 't Hooft coupling we find simple candidate gravity dual descriptions for both sets of vacua.

}
\begin{document}

%%%%%%%%%%%%%%%%%%%%%%%%%%%%%%%%%%%%%%%%%%%%%%%%%%%%%%
\section{Introduction}
%%%%%%%%%%%%%%%%%%%%%%%%%%%%%%%%%%%%%%%%%%%%%%%%%%%%%%

Quantum field theories which exhibit dynamical breaking of supersymmetry (DSB)
may be relevant in the description of Nature at the electroweak scale
\cite{Witten:1981nf}.  While theories which accomplish
DSB were found already in the early 1980s by
Affleck, Dine and Seiberg (for a review of early work,
see \cite{Affleck:1984xz}), the subject has retained much of its interest
over the past 25 years.
For instance, in the earliest examples, the supersymmetry breaking vacua
were global minima in chiral gauge theories.   However, it was realized
that by relaxing both of these criteria, one might obtain simpler 
examples
\cite{Dimopoulos:1997ww} which could yield less contrived realistic models
of gauge mediation \cite{gauge}.  This insight was further developed in many
papers \cite{models}, with perhaps the simplest idea
that can yield complete models appearing very recently in \cite{Dine:2006gm}.
Meta-stable supersymmetry breaking has also played a crucial r\^ole in many
recent constructions of string vacua \cite{Silverstein:2001xn,
Kachru:2003aw}, which quite plausibly realize
the idea of a ``discretuum'' proposed in
\cite{Bousso:2000xa}.

In the past decade, two new tools -- Seiberg duality \cite{Seiberg:1994pq}
and gauge/gravity duality \cite{Maldacena:1997re} -- have significantly
improved our ability to analyze the dynamics of
strongly coupled supersymmetric gauge
theories.  Since DSB is a strong-coupling phenomenon in many
instances, these new tools should be exploited to improve our understanding
of theories that exhibit DSB.

Gauge/gravity duality has already been applied in several different examples
to illuminate the physics of supersymmetry breaking
\cite{Maldacena:2001pb,Kachru:2002gs,
Berenstein:2005xa,Franco:2005zu,Bertolini:2005di}.  In one of these
instances, the 3d gauge theory analyzed in \cite{Maldacena:2001pb},
the direct field theory analysis \cite{Witten:1999ds} 
and the gravity analysis were
seen to agree.\footnote{The same gauge theory has been investigated in \cite{Schvellinger:2001ib}, 
without a focus on SUSY breaking.} In some other examples, such as those investigated in
\cite{Berenstein:2005xa,Franco:2005zu,Bertolini:2005di}, the dual
gauge theory does not 
admit any stable vacuum \cite{Franco:2005zu,Intriligator:2005aw,Brini:2006ej}. 
This is quite plausibly true in the large 't Hooft coupling gravity
dual as well, though compactifications of the scenario may fix this
problem \cite{Florea:2006si}, through 
the generation of baryonic couplings which were shown
to lead to stable non-supersymmetric vacua in some cases \cite{Argurio:2006ew}. In \cite{Franco:2006es}, it was
shown that these theories exhibit 
meta-stable non-supersymmetric vacua when massive flavors are added
by means of D7-branes.

The examples of \cite{Kachru:2002gs} (KPV) will be more relevant to our story.
That work builds directly on the beautiful paper of Klebanov and
Strassler \cite{Klebanov:2000hb}, where a smooth gravity dual was
found for the cascading $SU(N+M) \times SU(N)$ gauge theory of
branes and fractional branes at the conifold.
At the end of the cascade (for $N$ a multiple of $M$)
one finds a deformed conifold geometry
with a large sphere of radius $\sqrt{g_s M}$, and $M$ units
of RR three-form flux piercing the sphere.  It was proposed in
\cite{Kachru:2002gs} that by adding $p \ll M$ anti-D3 brane probes
to this system, one could obtain non-supersymmetric states in the
$SU(N+M-p) \times SU(N-p)$ supersymmetric gauge theory realized
by branes at the conifold.  Because the anti-D3 branes are attracted
to the warped tip of the geometry, the supersymmetry breaking states
have exponentially small vacuum energy.  In addition, they are connected
by finite energy bubbles of false vacuum decay, to the supersymmetric
vacua of the $SU(N+M-p) \times SU(N-p)$ theory \cite{Kachru:2002gs}.
This, together with
the fact that the boundary conditions at infinity in the gravity dual
are the ${\it same}$ for the supersymmetric and non-supersymmetric states
(in contrast to the situation described in \cite{Bena:2006rg}),
indicates that these are best thought of as dynamical supersymmetry
breaking states in the $SU(N+M-p) \times SU(N-p)$ gauge theory at
large 't Hooft coupling.
These states have played an important role in the KKLT construction
\cite{Kachru:2003aw}, and
in some models of inflation in string theory
\cite{Kachru:2003sx,DeWolfe:2004qx}.
More importantly for our purpose, it is obvious that the same analysis
would yield meta-stable KPV-like states in many other confining gauge
theories with smooth gravity duals.

In an a priori un-related development, it was recently found
in the elegant paper of
Intriligator, Seiberg and Shih \cite{Intriligator:2006dd} (ISS) that
even the simplest non-chiral gauge theories can exhibit
meta-stable vacua with DSB.
A
straightforward application of Seiberg duality to supersymmetric
$SU(N_c)$ QCD with $N_f$ slightly massive quark flavors of
mass $m \ll \Lambda_{QCD}$,
in the range $N_c+1 \leq N_f < {3\over 2} N_c$, yields
a dual magnetic theory which breaks supersymmetry at tree-level. \footnote{In \cite{Intriligator:2006dd}, it has 
been suggested that a meta-stable, supersymmetry breaking
vacuum also exists for $N_f=N_c$. We discuss the difficulties in the analysis of this
case in \S4.}
The supersymmetry
breaking vacuum is a miracle from the perspective of the electric
description, occurring in the strong-coupling regime of small field
VEVs where only the Seiberg dual description allows one to analyze
the dynamics.
And again, as will be important for us,
the analysis of the original paper can be extended to
provide many other examples.\footnote{
Some references which are in some respect relevant to our work are
\cite{Ooguri:2006pj,Banks:2006ma,Forste:2006zc,Amariti:2006vk}.}

On closer inspection, there are several qualitative similarities
between the KPV states (which were found using gauge/gravity duality)
and the ISS states (which were found using Seiberg duality).
In both examples, the supersymmetry breaking state
is related to the existence of a baryonic branch; in both examples, it is a
meta-stable state in a non-chiral gauge theory;
and in both examples, there is an intricate moduli space of Goldstone modes
(geometrized in the KPV case as the translation
modes of the anti-D3s at the end of the Klebanov-Strassler throat).
It is natural to wonder -- is there some direct relation between these
two classes of meta-stable states?\footnote{This question has been raised
by the present authors, H. Ooguri, N. Seiberg, H. Verlinde and many others.}

In this paper, we propose that at least in some cases, the answer is
${\it yes}$.  We analyze the gauge theory on D-branes at a certain simple
singularity (obtained from a $\IZ_2$ quotient of the conifold).
We find that this non-chiral gauge theory admits both supersymmetric and
supersymmetry breaking vacua -- the non-supersymmetric vacua being found 
by a simple generalization of ISS.
At large 't Hooft coupling, our system has a simple dual gravitational
description, along the lines of \cite{Klebanov:2000hb}.
We analyze the dual geometry, and propose natural candidates for
the gravity description of both sets of vacua -- with the non-supersymmetric
vacua arising from a simple generalization of KPV.
Our result suggests that there may be a general connection between the
classes of states that are unveiled using the two techniques.
Of course, this is example dependent: there is no firm argument that
meta-stable states which are present at large 't Hooft coupling continue
smoothly back to weak gauge coupling, or vice versa, in general.
For this reason, it is not clear that the states of the original ISS
model should have a simple gravity dual (or, that the states of the
original KPV model, should have a simple weak-coupling description).

The organization of our paper is as follows.  In \S2, we
describe the gauge theory that we will analyze, and a possible
brane configuration in string theory that gives rise to it.
In \S3, we will describe a duality cascade (similar to the one
in \cite{Klebanov:2000hb}) which will serve as a possible field theory
UV-completion of our model; the gauge theory we will focus on
arises at the end of the cascade.  In \S4, we argue, using field
theory techniques, that our gauge theory admits meta-stable states
that dynamically break supersymmetry, in addition to various
supersymmetric vacua.  In \S5, we provide a description of candidate 
gravity duals for both the supersymmetric and supersymmetry breaking
vacua.  We conclude with a discussion of possible directions for further
research and open questions in \S6. We have relegated 
a detailed discussion of the geometry of the moduli space, complex deformations 
from toric geometry and a (T-dual) IIA brane system, to the appendices.

%%%%%%%%%%%%%%%%%%%%%%%%%%%%%%%%%%%%%%%%%%%%%%%%%%%%%%
\section{The theory}
%%%%%%%%%%%%%%%%%%%%%%%%%%%%%%%%%%%%%%%%%%%%%%%%%%%%%%

In this section we present our model. We first present a prototype model
that has some of the properties we are interested in finding.
More precisely, this model exhibits ISS-like
meta-stable vacua in a gauge theory where the very small
flavor masses are not put in by hand, 
but instead are generated dynamically. Then, in order to illustrate
our ideas, we provide a concrete and simple D-brane/string
background which will serve as our privileged toy model in the
rest of the paper. It should be kept in mind, however, that our proposal
has a wide range of applicability and in principle many different
string
constructions could lead to this kind of dynamics.

%=====================================================
\subsection{Masses from quantum moduli spaces}
%=====================================================

\label{section_masses_from_mesons}

A key ingredient in the construction of ISS-like models
with non-supersymmetric meta-stable vacua is the existence of massive flavors.
Since we do not want to introduce external masses, the only remaining option
is to generate them dynamically by giving VEVs to fields participating
in a cubic (or higher) vertex in the superpotential. 
Moreover, if we want to forbid
those VEVs from relaxing to zero (which they may wish to do dynamically, since
the vacuum energy of the would-be non-supersymmetric vacua is proportional
to the quark masses), we have to impose at least one constraint
on them. The most natural such constraint is the one describing the
quantum deformed moduli space of $N_f=N_c$ SQCD 
\cite{Seiberg:1994bz}.\footnote{Indeed, the quantum
deformation was used for different but similar purposes in
\cite{Dimopoulos:1997ww}.} 
Hence, we are naturally
led to a model of a quiver gauge theory which features at least one
node in this regime.

\begin{figure}[ht]
  \centering
  \includegraphics[width=7cm]{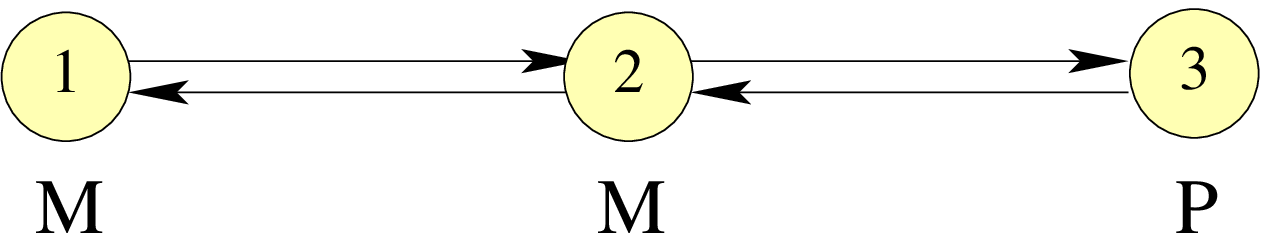}
  \caption{Sub-quiver that dynamically generates masses for the flavors of the third node.}
  \label{3_node_quiver}
\end{figure}

Before presenting all the details in later sections,
we briefly review here the simple mechanism that we want to propose in
order to generate such masses dynamically.
Let us consider the quiver diagram shown in \fref{3_node_quiver} with a tree level superpotential
involving at least the quartic interaction

\beq
W = \ldots + X_{21} X_{12}X_{23}X_{32}  + \ldots
\label{W_1}
\eeq
where here and henceforth, traces on gauge indices are understood.

Let us remind the reader that a quiver diagram is just a convenient pictorial way to encode the matter
content of certain gauge theories. Every node in the quiver represents a gauge group (in this paper we
will focus on $SU(N_i)$ gauge groups). An arrow connecting two nodes represents a chiral multiplet
transforming in the fundamental representation of the node at its tail and the anti-fundamental
representation of the node at its head.

%%%%%%%%%%%%%%%%%%%%%%%%%%%%%%%
%%%%%%%%%%%%%%%%%%%%%%%%%%%%%%%

Let us first suppose that the dynamical scales of the three gauge groups are
$\Lambda_1 \gg \Lambda_3 \gg \Lambda_2$. The strong dynamics of node 1 is described
in terms of mesons ${\cal M}^{i_2,j_2}=X^{i_2,i_1}_{21}X^{i_1,j_2}_{12}$
and baryons ${\cal B}=[X_{12}]^M$ and $\tilde{{\cal B}}=[X_{21}]^M$. The meson
${\cal M}$ transforms in the adjoint representation of node 2. The tree level
superpotential \eref{W_1} becomes
\beq
W=  \ldots + {\cal M}X_{23}X_{32}+\ldots~.
\label{W_2}
\eeq

Node 1 has $M$ colors
and $M$ flavors and hence leads to a quantum modified moduli space, corresponding to
\beq
\det_{i_2,j_2} {\cal M}-{\cal B} \tilde{{\cal B}}=\Lambda_1^{2M}~.
\label{quantum_constraint}
\eeq

On the mesonic branch, $\langle {\cal B} \rangle = \langle \tilde{{\cal B}} \rangle=0$
and $\det_{i_2,j_2} \langle{\cal M}\rangle=\Lambda_1^{2M}$. The gauge group
is higgsed down to $SU(P)\times U(1)^{M-1}$, with the abelian factors
coming from the adjoint higgsing of the second node.

From the point of view of node 3, $X_{23}$ and $X_{32}$ give rise to $M$ flavors
with $i_2$ becoming a flavor index. When the expectation value of ${\cal M}$ is
plugged into \eref{W_2}, it gives rise to non-zero masses for these flavors. The theory
becomes precisely $SU(P)$ SQCD with $M$ massive flavors. The masses are constrained by
\eref{quantum_constraint} but are dynamical quantities. 
In order to follow the argument of ISS, we would then perform a Seiberg
duality on node 3, which is the one responsible for DSB.

The $\Lambda_1 \gg \Lambda_3 \gg \Lambda_2$ regime we have just discussed is the simplest to study, since 
as we lower the energy scale we first generate the masses for the flavors of node 3. However, 
in order for the analysis of ISS to be valid, flavor masses should be much smaller than the dynamical scale 
$\Lambda_3$.

Since the masses are dictated by $\Lambda_1$, it is more natural to achieve a small mass to $\Lambda_3$ ratio
for $\Lambda_3 \gg \Lambda_1 \gg \Lambda_2$. If $P<M$, we begin by dualizing node 3.  If $P+1 \leq M < \frac{3}{2} P$, node 3
is IR free in the dual theory. 
In the magnetic theory, there are mesons ${\cal N}^{i_2,j_2}=X_{23}^{i_2,i_3}X_{32}^{i_3,j_2}$ and
``dual quarks'' $Y_{32}$ and $Y_{23}$. The superpotential becomes

\beq
W_{mag} = \ldots + X_{21} X_{12}~{\cal N}+{\cal N}~Y_{23} Y_{32} + \ldots ~,
\label{W_3}
\eeq
where the first terms comes from \eref{W_2} and the second one is the usual cubic coupling between Seiberg mesons
and dual quarks. At the $\Lambda_1$ scale, node 1 develops a quantum moduli space as in \eref{quantum_constraint}.  
Going to the mesonic branch, the superpotential becomes

\beq
W_{mag} = \ldots + \langle {\cal M} \rangle ~{\cal N}+{\cal N}~ Y_{23} Y_{32} + \ldots ~.
\label{W_4}
\eeq
The rank condition of ISS \cite{Intriligator:2006dd} arises from the F-term of ${\cal N}$. 
\footnote{The rank condition corresponds to the inability to have all $F_{{\cal N}^{i_2,j_2}}=0$
because the rank of node 3 in the magnetic theory (which determines the maximum possible rank of $Y_{23} Y_{32}$) 
is $(M-P)$, smaller than the dimension of ${\cal M}$.}

We thus conclude that our model potentially admits meta-stable
nonsupersymmetric vacua for both $\Lambda_1 \gg \Lambda_3 \gg \Lambda_2$ and 
$\Lambda_3 \gg \Lambda_1 \gg \Lambda_2$. 
In fact, we have shown that going on the quantum deformed mesonic branch 
of node 1, and performing the Seiberg duality on node 3, are commuting
operations. Hence we do not have to assume an a priori hierarchy between
the scales $\Lambda_1$ and $\Lambda_3$.
A more detailed examination in \S4 will show 
that the only really plausible case is when 
$M=P$.\footnote{The argument for the existence
of a meta-stable vacuum when $P=M$ is more subtle, see \S4.}
There, it is natural to assume
strong dynamics at both nodes 1 and 3 simultaneously.
In \S\ref{section_VY} we argue that essentially any hierarchy 
between $\Lambda_1$ and $\Lambda_3$ is attainable in a string theory realization of this model.

Since our goal is to generate non-zero masses dynamically, it is crucial to ensure that
they are stable against relaxation to zero or infinity. The first scenario could occur
if there is an instability towards condensation of baryons, and would destroy the possibility of 
SUSY breaking. Settling this question is difficult, since the SQCD node is in a confining regime and
non-computable corrections to the K\"ahler potential are present. There is no obvious sign of an instability
in the gravity dual that we construct later. In the rest of the paper, we will work under the assumption 
that the mesonic branch of node 1 is stable. This is a question that certainly deserves further study. 
An alternative direction would be to investigate the issue of stability of dynamical masses in similar theories for which it
is possible to work in the free-magnetic regime. 

The gauge theory described above can be viewed as a sub-sector embedded in a larger quiver,
with possibly more gauge groups, fields (even charged under node 3) and superpotential interactions.

%=====================================================
\subsection{A $\IZ_2$ orbifold of the conifold}
%=====================================================

\label{section_orbifold_conifold}

In this section we present a theory that contains the sub-quiver discussed
in \S\ref{section_masses_from_mesons} and hence has the appropriate non-chiral
matter and quartic interactions to generate dynamical masses by quantum deformation
of the moduli space. In addition, this model has a concrete string theory realization
in terms of D-branes probing a singularity.

\begin{figure}[ht]
  \centering
  \includegraphics[width=4cm]{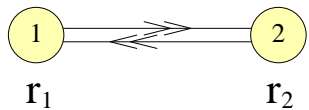}
  \caption{Quiver diagram for the conifold with $SU(r_1) \times SU(r_2)$ gauge group.}
  \label{quiver_conifold}
\end{figure}

The model we consider is a non-chiral $\IZ_2$ orbifold of the conifold
(see e.g. \cite{Uranga:1998vf}). It follows from the
conifold gauge theory by the standard orbifolding procedure. \fref{quiver_conifold} shows the conifold
quiver for arbitrary ranks $r_1$ and $r_2$. The corresponding superpotential is
\beq
W=h \epsilon^{ij} \epsilon^{kl} A_i B_k A_j B_l~.
\label{W_conifold}
\eeq

%%%%%%%%%%%%%%%%%%%%%%%%%%%%%%%
%%%%%%%%%%%%%%%%%%%%%%%%%%%%%%%

The conifold may be described by the following equation in four complex variables
\beq
xy=zw~.
\label{conifold}
\eeq

The gauge invariant variables are given in terms of the chiral bifundamental fields by
\beq
x \sim A_1 B_1 \ \ \ \ \ y \sim A_2 B_2 \ \ \ \ \ z \sim A_1 B_2 \ \ \ \ \ w \sim A_2 B_1~.
\label{variables}
\eeq

The $\IZ_2$ orbifold group has a single generator $\theta$, whose geometric action corresponds to
quotenting by the following identifications \footnote{Other quotients are possible. For example $x \rightarrow -x$,
$y \rightarrow -y$, $z \rightarrow -z$ and $w \rightarrow -w$ gives 
rise to the also well studied cone over the zeroth Hirzebruch
surface $F_0$.}
\beq
\begin{array}{lcl}
x \rightarrow x  & \ \ \ \ \ & y \rightarrow y \\
z \rightarrow -z  & \ \ \ \ \ & w \rightarrow -w~.
\end{array}
\label{quotient}
\eeq

We can define new variables that are invariant under the orbifold group. From the action in \eref{quotient},
they are $x'=x$, $y'=y$, $z'=z^2$ and $w'=w^2$. It is now straightforward to see that the orbifolded geometry
is given by the following equation
\beq
x'^2 y'^2=w'z'~.
\label{equation_Z2_conifold}
\eeq

We re-derive this equation from the field theory in appendix \ref{appendix_geometry}. The resulting geometry turns
out to be toric. A convenient way to describe toric geometries is via toric diagrams and $(p,q)$ webs, this being a useful
way for visualizing the geometric structure of the singularity, the possible complex structure deformations, the number (and type)
of allowed fractional branes, etc. Appendix \ref{appendix_toric} provides a brief review of toric
geometry and the applications of $(p,q)$ webs together with a more detailed analysis of the case under investigation.

Let us now construct the orbifolded gauge theory. We start from the conifold theory with $SU(r_1) \times SU(r_2)$
gauge group.
The coordinates are related to the chiral fields as in \eref{variables}. It follows that
the geometric action of the orbifold group generator $\theta$ given by \eref{quotient}
can be implemented by
\beq
\begin{array}{lcl}
A_1 \rightarrow -A_1  & \ \ \ \ \ & A_2 \rightarrow A_2 \\
B_1 \rightarrow -B_1  & \ \ \ \ \ & B_2 \rightarrow B_2~.
\end{array}
\eeq

\begin{figure}[ht]
  \centering
  \includegraphics[width=5cm]{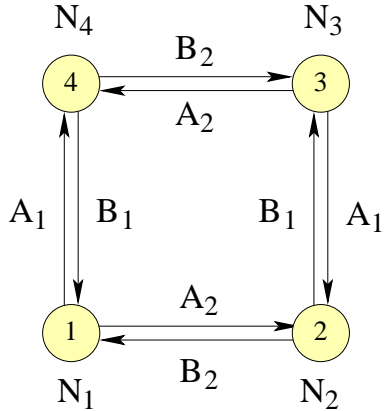}
  \caption{Quiver diagram for the $\IZ_2$ orbifold of the conifold under consideration,
 for arbitrary numbers of fractional and regular
D3-branes. We have labeled bifundamentals according to the parent field.}
  \label{quiver_Z2_conifold_labeled}
\end{figure}

In addition, we have to specify the action of $\theta$ on the Chan-Paton factors of the two gauge
groups. We take it to be
\beq
\begin{array}{l}
\gamma_{\theta,1}=diag(\id_{N_1},-\id_{N_3}) \\
\gamma_{\theta,2}=diag(\id_{N_2},-\id_{N_4})
\end{array}
\eeq
where $N_1+N_3=r_1$ and $N_2+N_4=r_2$. The resulting gauge group
is $\prod_{i=1}^4 SU(N_i)$. Both the conifold gauge theory and the orbifold we are studying
are completly non-chiral. Hence there are no constraints
on the ranks of the various gauge groups coming from 
anomaly cancellation --  
they are completely arbitrary.
Combining the geometric and Chan-Paton
actions, we conclude that the original fields give rise to the following
chiral multiplets
\beq
\begin{array}{lcl}
A_1 \rightarrow X_{14},X_{32} & \ \ \ \ \ & A_2 \rightarrow X_{12},X_{34} \\
B_1 \rightarrow X_{41},X_{23} & \ \ \ \ \ & B_2 \rightarrow X_{21},X_{43}~.
\end{array}
\eeq
The resulting quiver is shown in \fref{quiver_Z2_conifold_labeled}, in which
we have indicated the parent fields.

%%%%%%%%%%%%%%%%%%%%%%%%%%%%%%%
%%%%%%%%%%%%%%%%%%%%%%%%%%%%%%%

The superpotential follows from projecting \eref{W_conifold} and is given by
\beq
W=h\left( X_{14}X_{41}X_{12}X_{21}+X_{32}X_{23}X_{34}X_{43}-X_{14}X_{43}X_{34}X_{41}-X_{32}X_{21}X_{12}X_{23} \right)~.
\label{W_Z2_conifold}
\eeq

%=====================================================
\subsection{Fractional branes}
%=====================================================

This theory has three types of fractional branes. They correspond to different ways in which D5-branes
can be wrapped over 2-cycles collapsed at 
the tip of the singularity. At the level of the gauge theory,
this corresponds to the already noted fact that anomaly freedom does
not constrain the ranks of the four gauge groups. 
A convenient basis of fractional branes is given by the rank
vectors $(1,1,0,0)$, $(0,0,1,0)$ and $(1,0,0,0)$. In this language, $(1,1,1,1)$ represents a regular D3-brane.

In \cite{Franco:2005zu}, a classification of fractional branes based on the IR behavior they trigger was
introduced. It turns out there are three different classes of fractional branes:

\bigskip

\noindent $\bullet$ {\it ${\cal N}=2$ fractional branes:} the quiver gauge theory on them (in the absence
of regular D3-branes) has closed oriented loops (gauge invariant operators) that do not appear in the superpotential.
Hence, these fractional branes have flat directions parametrized by the expectation values of these mesonic operators.
From a geometric point of view, these fractional branes arise when the singularities are not isolated, but have curves
of $\IC^2/\IZ_N$ singularities passing through them. The IR dynamics of the gauge theories (instantons and
Seiberg-Witten points) map to 
an enhan\c{c}on mechanism in a gravity dual description \cite{Johnson:1999qt}.

\bigskip

\noindent $\bullet$ {\it Deformation fractional branes:} the quiver on these branes
is either a set of decoupled nodes, or nodes with closed loops that appear in the superpotential.
The ranks of all gauge factors are equal.
Geometrically, these fractional branes are associated with a possible
complex deformation of the singularity. In the gauge theory, the
gauge groups which are involved undergo confinement. 
This is translated to a complex
structure deformation leading to finite size 3-cycles in the gravity dual.

\bigskip

\noindent $\bullet$ {\it DSB fractional branes:}
these are fractional branes of any other kind, hence they provide the generic case. In this case, the
non-trivial gauge factors have unequal ranks. Geometrically, they are associated with geometries for which
the corresponding complex deformation is obstructed. The gauge theory dynamics corresponds to the appearance
of an Affleck-Dine-Seiberg (ADS) superpotential \cite{Affleck:1983mk} 
that removes the 
supersymmetric vacuum \cite{Berenstein:2005xa,Franco:2005zu,Bertolini:2005di}.
Furthermore (as first discussed in \cite{Franco:2005zu} 
and later studied in detail in
\cite{Intriligator:2005aw,Brini:2006ej}) 
the gauge theory has a runaway behavior towards infinity
parametrized by di-baryonic operators.

\bigskip

It is important to keep in mind that combining fractional branes in one or more of these classes can lead to
fractional branes of another kind.
In the example under study, the $(0,0,1,0)$ and $(1,0,0,0)$ branes are deformation fractional branes, while $(1,1,0,0)$ is
an ${\cal N}=2$ fractional brane.

\begin{figure}[ht]
  \centering
  \includegraphics[width=7cm]{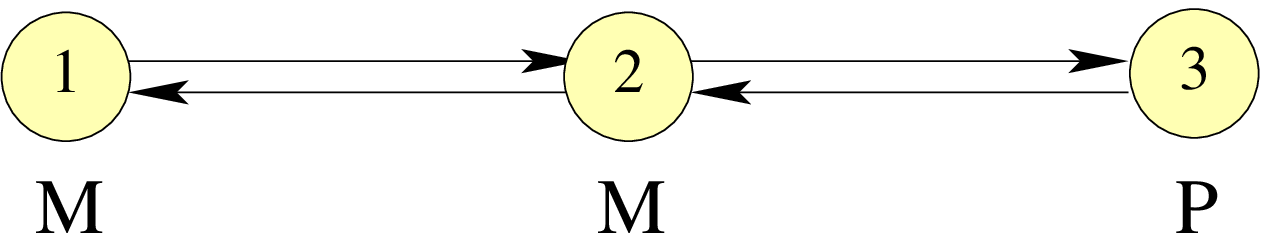}
  \caption{Quiver diagram for $(M,M,P,0)$ ranks.}
  \label{quiver_Z2_conifold_2}
\end{figure}

In order to make contact with our discussion in \S\ref{section_masses_from_mesons}, we will focus on the quiver
with ranks $(M,M,P,0)$. It can be engineered by using $M$ $(1,1,0,0)$ and $P$ $(0,0,1,0)$ fractional branes.
The resulting quiver is shown in \fref{quiver_Z2_conifold_2}. The surviving superpotential is given by
\beq
W=h X_{32}X_{21}X_{12}X_{23}
\label{W_Z2_conifold_2}
\eeq
where, for later convenience, we have changed the coupling $h$ by an overall sign with respect to \eref{W_Z2_conifold}.

%%%%%%%%%%%%%%%%%%%%%%%%%%%%%%%
%%%%%%%%%%%%%%%%%%%%%%%%%%%%%%%

This is precisely the theory discussed in \S\ref{section_masses_from_mesons}.
In the next section we explain in detail how this theory arises at the 
IR end of a duality cascade.  Its 
vacuum structure will be analyzed in \S\ref{section_meta-stable_vacuum}.

%%%%%%%%%%%%%%%%%%%%%%%%%%%%%%%%%%%%%%%%%%%%%%%%%%%%%%
\section{The duality cascade}
%%%%%%%%%%%%%%%%%%%%%%%%%%%%%%%%%%%%%%%%%%%%%%%%%%%%%%

It is always the case that supergravity solutions describing the geometry generated by sets of fractional branes
have (logarithmically) running fluxes. The cut-off one is forced to introduce in order to regulate the logarithms
automatically brings regular branes into the story,
 from which the fractional branes cannot then be disentangled, if one
aims for a weakly curved supergravity description. The same 
holds for the quiver in \fref{quiver_Z2_conifold_2}
which should be thought of 
as part of a more general theory, involving regular branes, too.

In general, the dynamics of stacks of fractional and regular branes gets a natural interpretation in terms of a
duality cascade (for ${\cal N}=1$ gauge/gravity dualities). This is well understood for the case of the conifold
\cite{Klebanov:2000nc,Klebanov:2000hb,Strassler:2005qs}. However, when departing from this well known example and focusing
on more involved theories, it is not straightforward to visualize a specific pattern
\cite{Franco:2004jz,Herzog:2004tr}. Here, we will provide one.
The discussion is somewhat involved, and a reader who is interested purely
in the dynamics of the quiver in Figure 4 can skip this section on a first
reading.

\begin{figure}[ht]
  \centering
  \includegraphics[width=4.4cm]{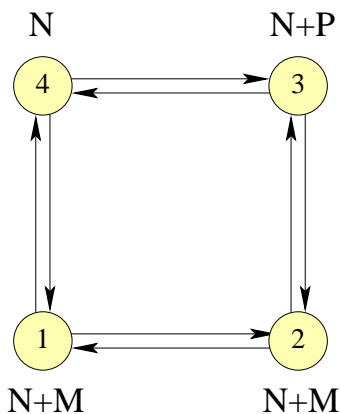}
  \caption{Quiver generated by $N$ regular branes, $M~(1,1,0,0)$ and $P~(0,0,1,0)$ fractional branes.}
  \label{uv}
\end{figure}

In general, the physics of a cascade is obtained when one perturbs a fixed point of some SCFT, generated
by $N$ regular D3-branes at a singularity, with some (smaller) number of fractional D3-branes. This brings
the theory out of the fixed point and triggers a non-trivial RG-flow. What happens, in a quite model-independent way,
is that the number of regular branes effectively diminishes along the flow and the IR dynamics of the theory
is determined by fractional branes only. Therefore, the natural guess for the UV theory generating via RG-flow the
dynamics depicted in \fref{quiver_Z2_conifold_2} in the IR, should be one described by $N$ regular branes,
$M$ ${\cal N}=2$ fractional branes and $P$ deformation fractional branes corresponding to the quiver in \fref{uv},
%%%%%%%%%%%%%%%%%%%%%%%%%%%%%%%
%%%%%%%%%%%%%%%%%%%%%%%%%%%%%%%
with in addition the tree-level superpotential 
\be
\label{wtree}
W = X_{14}X_{41}X_{12}X_{21}+X_{32}X_{23}X_{34}X_{43}-X_{14}X_{43}X_{34}X_{41}-X_{32}X_{21}X_{12}X_{23}
\ee
which is the same as (\ref{W_Z2_conifold}) up to an overall constant\footnote{In this superpotential and subsequent ones in 
this section we omit coupling constants. At every Seiberg duality
dynamical scales must be matched and mesonic fields are normalized in order to give them canonical dimensions
according to standard rules. For the sake of simplicity, we choose the simplest approach of reinserting the 
superpotential coupling in the final result \eref{final_W}.}. In 
what follows, we show that this is a correct guess.

\begin{figure}[ht]
  \centering
  \includegraphics[width=4.7cm]{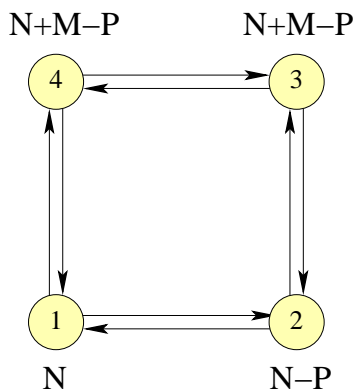}
  \caption{The theory after one full cascade step.}
  \label{first}
\end{figure}

For a cascade to actually occur, the theory should be self-similar, that is
after a certain number of Seiberg dualities it should return to
itself (including the superpotential), but with a reduced number $N$ of regular
branes. This model, unlike its cousin the conifold,
needs more than one Seiberg duality to display its
self-similar structure. In fact, a possible pattern is via four
subsequent duality steps, where these are taken on the 
gauge groups of nodes 1, 3, 2 and
4, respectively. With properly chosen initial conditions one can
show that the theory one obtains after this pattern of Seiberg dualities
(a {\it single} cascade step) is the one depicted in the quiver in \fref{first} with superpotential \eref{wtree}.
\footnote{In the process a number of non-trivial 
intermediate steps occur, to make
the theory self-similar. In particular, there are extra (adjoint)
fields generated at each duality step, together with 
the corresponding cubic superpotential terms.
Every two Seiberg duality steps, the adjoint fields pair up 
in mass terms and can be integrated
out consistently.}

%%%%%%%%%%%%%%%%%%%%%%%%%%%%%%%
%%%%%%%%%%%%%%%%%%%%%%%%%%%%%%%

This is the same as the original theory, after taking $N \rightarrow N-P$
and re-labeling the indices as $1 \leftrightarrow 3~,~2 \leftrightarrow 4$.
This proves the self-similar structure of the theory and the existence of
the cascade. Notice that only deformation branes enter the cascade:
the diminishing of the number of regular branes along the cascade is only affected
by the $P$ deformation branes. 
It may be worth noticing,
at this point, that the several (four in the present case) Seiberg dualities needed
to recover the self-similarity of the theory could be thought of as if they were done simultaneously.
What is more physical, and indeed is usually captured by the gravity dual through the running of
the $B_2$ (and consequently $F_5$) flux, are the {\it cascade} steps.

Suppose now $N=kP$. Naively, one would say that after $k$ cascade steps of the type discussed
above, $N$ gets reduced to $0$ yielding the theory we are looking for, 
namely the quiver in Figure \ref{quiver_Z2_conifold_2}.
This turns out to be correct. However, one should remember that strictly speaking the final step of a cascade is
not described by a Seiberg duality since generically, at such energy scales, at least one gauge group ends up having
$N_f=N_c$ and the moduli space is deformed at the quantum level. 
As in the well-studied case of the conifold,
in our case one can show 
that the strongly coupled gauge group confines and along the baryonic branch
one indeed ends up with the theory in Figure \ref{quiver_Z2_conifold_2} 
(while, as in the conifold case, the
mesonic branch has instead $P$ 
freely-moving regular branes in the background
\cite{Dymarsky:2005xt}).

%=====================================================
\subsection{The last cascade step}
%=====================================================

\begin{figure}[ht]
  \centering
  \includegraphics[width=4.4cm]{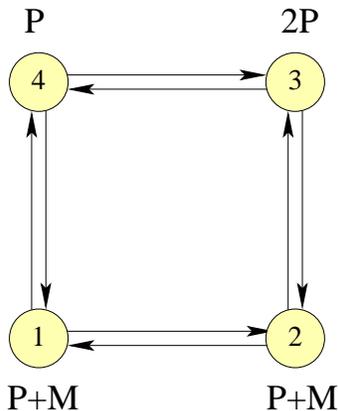}
  \caption{The theory at the next-to-last step of the cascade, $N=P$.}
  \label{next}
\end{figure}

Let us treat the last cascade step more carefully (this is done along the lines of \cite{Bertolini:2005di}).
After $k-1$ steps one gets effectively $N=P$ with the corresponding quiver reported in Figure \ref{next}.

%%%%%%%%%%%%%%%%%%%%%%%%%%%%%%%
%%%%%%%%%%%%%%%%%%%%%%%%%%%%%%%

\begin{figure}[ht]
  \centering
  \includegraphics[width=4.4cm]{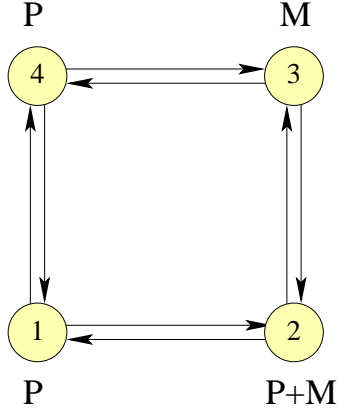}
  \caption{The intermediate step.}
  \label{next2}
\end{figure}

Apparently, this looks different from known examples in the literature since, here, for any node $N_f>N_c$ and the moduli
space is {\it not} deformed at the quantum level. However, after taking the first two Seiberg dualities,
on nodes 1 and 3, respectively, one gets
\be
N_{1}' = 2P+M - P- M = P~~\mbox{and}~~N_{3}' = 2P+ M -2P = M
\ee
with corresponding quiver as in \fref{next2}
%%%%%%%%%%%%%%%%%%%%%%%%%%%%%%%
%%%%%%%%%%%%%%%%%%%%%%%%%%%%%%%
and superpotential (\ref{wtree}). As already noticed, 
the adjoint fields generated by Seiberg duality
are massive and can be integrated out, together 
with their superpotential interaction terms. Here the
process stops since node 2 
now has $N_c=N_f=P+M$ and the moduli space gets deformed at the quantum level. 
Let us then focus on the dynamics of node 2. The quantum constraint reads
\be
\mbox{det}{\cal M} - {\cal B}\tilde{\cal B} = \Lambda_2^{2(P+M)}~.
\label{quantum_1}
\ee

\begin{figure}[ht]
  \centering
  \includegraphics[width=4.4cm]{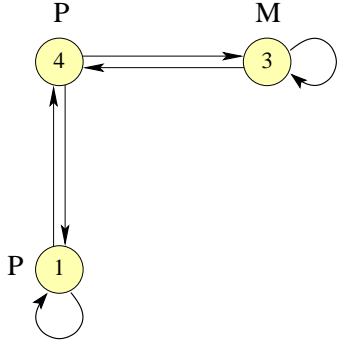}
  \caption{The theory below the $\Lambda_2$ scale.}
  \label{bottom2}
\end{figure}

The baryons can be written as
\be
{\cal B} \sim (X_{32})^M (X_{12})^P~~~,~~~\tilde{\cal B} \sim (X_{23})^M (X_{21})^P
\ee
where contraction with the Levi-Civita tensor is understood. The meson matrix can be written as
\be
\label{meson}
{\cal M} = \left(\matrix{X_{12} X_{21}& X_{12} X_{23} \cr X_{32} X_{21} & X_{32} X_{23} \cr}
\right) \equiv
\left(\matrix{{\cal M}_{11}&{\cal M}_{13} \cr {\cal M}_{31}&{\cal M}_{33}\cr}\right)~.
\ee
The superpotential is given by
\be
\label{wquantum}
W =  X_{14} X_{41} {\cal M}_{11}  + {\cal M}_{33} X_{34}X_{43} - X_{14}X_{43}X_{34}X_{41}-  {\cal M}_{31} {\cal M}_{13} 
+ \lambda \left(\mbox{det}{\cal M} - {\cal B}\tilde{\cal B} - \Lambda_2^{2(P+M)} \right)~
\ee
where $\lambda$ is a gauge invariant chiral field that acts as a Lagrange multiplier enforcing \eref{quantum_1}.
Let us focus on the baryonic branch, $\mbox{det}{\cal M} = \lambda = 0~,~ {\cal B}\tilde{\cal B}
= \Lambda_2^{2(P+M)}$. Looking at the superpotential (\ref{wquantum}) one sees both ${\cal M}_{13}$ and
${\cal M}_{31}$ are massive and can be integrated out. Node 2 confines and below the strong coupling scale
$\Lambda_2$ the theory gets reduced to the quiver in Figure \ref{bottom2},
%%%%%%%%%%%%%%%%%%%%%%%%%%%%%%%
%%%%%%%%%%%%%%%%%%%%%%%%%%%%%%%
with superpotential
\be
W = X_{14} X_{41} {\cal M}_{11} + {\cal M}_{33} X_{34}X_{43} - X_{14}X_{43}X_{34}X_{41}~.
\ee

\begin{figure}[ht]
  \centering
  \includegraphics[width=4.4cm]{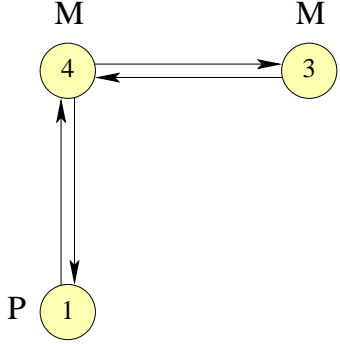}
  \caption{The theory at the bottom of the cascade.}
  \label{bottom3}
\end{figure}

Now we can finally Seiberg dualize on node 4, which has $N_f=M+P > N_c = P$. After dualization its rank becomes $M$.
There will be a meson matrix ${\cal N}$ similar to (\ref{meson}) in terms of which the effective superpotential reads
\be
W = {\cal N}_{11}{\cal M}_{11} + {\cal M}_{33} {\cal N}_{33}  - {\cal N}_{13} {\cal N}_{31} - {\cal N}_{11} Y_{14} Y_{41} - 
{\cal N}_{33} Y_{34} Y_{43} + {\cal N}_{13} Y_{34} Y_{41} + {\cal N}_{31} Y_{14} Y_{43}
\ee
where the $Y_{ij}$ are the dual quarks. Now we can integrate out all the 
mesons getting the superpotential
\be
W =  h \, Y_{14} Y_{43} Y_{34} Y_{41}
\label{final_W}
\ee
where we have made the superpotential coupling explicit, calling it $h$ as in the original theory for simplicity. 
The corresponding quiver is shown in \fref{bottom3}.

%%%%%%%%%%%%%%%%%%%%%%%%%%%%%%%
%%%%%%%%%%%%%%%%%%%%%%%%%%%%%%%

This is nothing but the theory of Figure \ref{quiver_Z2_conifold_2} and superpotential \eref{W_Z2_conifold_2}, 
after a trivial re-labeling of letters ($Y \rightarrow X$) and indices ($3 \leftrightarrow 1 , 4 \leftrightarrow 2$),
as promised.

%%%%%%%%%%%%%%%%%%%%%%%%%%%%%%%%%%%%%%%%%%%%%%%%%%%%%%
\section{The meta-stable non-supersymmetric vacuum}
%%%%%%%%%%%%%%%%%%%%%%%%%%%%%%%%%%%%%%%%%%%%%%%%%%%%%%
\label{section_meta-stable_vacuum}

The theory we want to analyze has a rich structure and, as we are going to show, admits both supersymmetric as well
as (meta-stable) non-supersymmetric vacua. In this section we will be concerned with the latter and argue, in particular,
that they actually arise only in the case where $P=M$.

Let us first consider our 3-node quiver in Figure \ref{quiver_Z2_conifold_2}, 
with $P<M$. In this case, we have $N_f=M>P=N_c$. This is potentially 
interesting because we could in principle be
in the range $N_c+1\leq N_f < \frac{3}{2}N_c$. In this case the theory has an IR free Seiberg dual description and
the arguments of ISS in favor of meta-stability are conclusive.

On the baryonic branch of node 1, a standard analysis
shows that one has then to perform
a Seiberg duality on node 3, and integrate out two adjoints on node 2.
We eventually end up in a two node $SU(M)\times SU(M-P)$
quiver with vanishing tree level superpotential. 
What was formerly labeled node 2
develops an ADS runaway superpotential and there is no stable vacuum.

On the mesonic branch of node 1,
we have the supersymmetric moduli space of the ${\cal N}=2$ fractional branes.
This is best seen assuming that
the VEVs of the mesons ${\cal M}$ give a mass to the flavors of node 3, and
integrating the latter out. Hence at every point on the moduli space
we have the $P$ vacua of $SU(P)$ SYM.

As for the non supersymmetric states, 
following  \S\ref{section_masses_from_mesons} and the analysis of
\cite{Intriligator:2006dd}, we would expect a meta-stable vacuum with vacuum energy
\beq
V_{meta} \sim |\Lambda_3|^2 \sum_{i=1}^P |h m_i|^2
\eeq
where $m_1, \dots ,m_P$ are the $P$ eigenvalues of ${\cal M}$ with smallest absolute values.
Clearly, the constraint $\det {\cal M} = \Lambda_1^{2M}$ does not set a minimum value
for them. What happens is that $m_1, \dots ,m_P$ relax to zero, while some of the other eigenvalues run away to
infinity so that $\det {\cal M}$ remains constant. As a result, we conclude that there is no SUSY breaking ISS state
for $P<M$.

We can also briefly ask what happens in the case $P>M$. The baryonic branch can be shown to have a runaway because the $SU(P)$ node
has an ADS superpotential which drives the meson ${\cal M}$ to infinity.
The supersymmetric mesonic
branch is exactly as above. The ISS-like meta-stable states are absent
because $N_f<N_c$ for the $SU(P)$ gauge group.

We are thus left with only one potentially interesting case, namely $P=M$.
We now show that there is indeed evidence, albeit weaker, for both 
supersymmetric and supersymmetry breaking states in this case.

%=====================================================
\subsection{Meta-stable vacuum in $N_f=N_c$ SQCD}
%=====================================================
We begin by reviewing how we can heuristically recover the meta-stable
vacua conjectured to exist by Intriligator, Seiberg and Shih
\cite{Intriligator:2006dd} in $N_f=N_c$ SQCD.
We note that their conjecture for this case is based on somewhat weaker
evidence than for the cases with $N_c + 1 \leq N_f < {3\over 2} N_c$.

Supposing that the quark chiral fields have a small mass $m$, the low energy
effective superpotential for the mesons ${\cal M}=Q \tilde Q$ and the baryons is
\beq
W=\tr m ~{\cal M} + \lambda(\det {\cal M} - {\cal B} \tilde {\cal B} - \Lambda^{2N_c})~.
\eeq

The F-term conditions are
\beq
\det {\cal M} - {\cal B} \tilde {\cal B} = \Lambda^{2N_c}~, \qquad m +\lambda (\det {\cal M} ) {\cal M}^{-1}=0~,
\qquad \lambda {\cal B} = 0 =  \lambda \tilde {\cal B}~.
\eeq

If we are to satisfy the equation involving $m$, we must have $\lambda\neq 0$.
This implies that ${\cal B}=\tilde {\cal B}=0$, which we usually refer to as being on the
mesonic branch. Then the constraint $\det {\cal M}=\Lambda^{2N_c}$ eventually leads
to a SUSY vacuum with mesonic VEVs
\beq
{\cal M}=m^{-1} (\det m)^\frac{1}{N_c} \Lambda^{2} \sim \Lambda^{2}~.
\eeq
We assume here and below that the masses are all of the same order. With this
assumption, note that the mass dependence of the mesonic VEVs actually cancels.

Alternatively, if
we want to give non zero VEVs to the baryons, ${\cal B}, \tilde {\cal B} \neq 0$,
which is usually referred to as being on the baryonic branch, we must
set $\lambda=0$. The equation involving $m$ can no longer be satisfied, and
this gives rise to the non-vanishing F-terms contributing to the vacuum
energy. Taking for convenience a mass matrix proportional to the
identity, we would get
\beq
V_{meta} \sim N_c |m|^2 |\Lambda|^2~.
\eeq

The factor of $\Lambda$ comes from the proper normalization of the meson field
${\cal M}$ (whose F-term is non-vanishing), under the assumption
that we are at a smooth (though strongly coupled) point in the moduli space.

Note that though we have referred to mesonic and baryonic branches, they are
not really disconnected.\footnote{
We assume here that even on the non SUSY states,
the constraint has to be applied. Indeed, the
constraint is in fact a non dynamical F-term, hence it cannot be violated.
It is a relation in the chiral ring which has no classical
nor perturbative corrections in any state.
The only way it could be evaded is by non-perturbative
$\bar D$ exact contributions
which could be non vanishing in a non SUSY state. Though presumably most of
these corrections can be excluded by further considerations,
even if they are there, they will not change
significantly the structure of the constraint, just shifting by a small amount
the effective value of $\Lambda^{2N_c}$ in the non SUSY state.}
The only ($N_c$) SUSY vacua have fixed non zero VEVs for the mesons and
vanishing VEVs for the baryons. Thus {\em any} state with different VEVs
on the original quantum moduli space is a non SUSY state. Clearly
most of those states are not meta-stable (they have tadpoles),  and
smoothly relax to the SUSY
vacua. The argument by ISS, based on $m\ll \Lambda$ and the relation
to the more controlled
case of $N_f=N_c+1$, is that far from the SUSY states,
i.e. when the baryons have the largest allowed VEVs, the state
is meta-stable. It can be shown that the tree-level potential has indeed
an extremum there, with flat directions that are not lifted by loops of
massive but light flavors. We should emphasize again 
that the $N_f=N_c$ case is the one in which the
ISS arguments favoring the existence of a meta-stable vacuum are 
least explicit.

Meta-stability should of course eventually be verified by checking how the
pseudo-moduli are lifted around that vacuum by higher order (in $1/\Lambda^2$) terms in the
K\"ahler potential. Relevant considerations
about the K\"ahler potential in a system very similar to the one discussed here
and below
can be found in \cite{Dimopoulos:1997ww}, where arguments in favor of
a local minimum are given. It would also be interesting to investigate the
trajectories in field space connecting the meta-stable vacuum to the SUSY ones along which
the potential barrier coming from the tree-level superpotential is minimum.
The height and width of the potential barrier determine the
lifetime of the meta-stable vacuum.

Here, we make one more comment about the lifting of pseudo-moduli that
applies to compactifications of our solutions (along the lines of 
\cite{Giddings:2001yu}). 
On the baryonic branch, there is a $U(1)_B$ symmetry which is spontaneously
broken by the baryon VEV.  In the decoupled theory, this $U(1)_B$ is a 
global symmetry, and its breaking gives rise to a Goldstone boson and its
saxion partner (as discussed in, for instance, \cite{Gubser:2004qj}). 
The saxion is a dangerous direction -- its masslessness is not protected
by a symmetry, and in any non-supersymmetric vacuum, one can worry that
it could become tachyonic.  In {\it compactifications} of
this kind of theory, there is one known computable contribution to the mass
that acts in the direction of stabilizing the supersymmetry breaking 
vacuum.  

String compactifications do not have continuous global symmetries.  Instead, the
$U(1)_B$ becomes a gauge symmetry.  The baryon VEVs then give mass to
the U(1) gauge boson via the Higgs mechanism, and the would-be Goldstone
multiplet is ``eaten."  The relevant mass that is imparted is 
proportional to the product of the gauge coupling and the baryon VEVs,
and following \S6\ of \cite{Gubser:2004qj}, this can be estimated as follows.
If the hierarchy of scales generated by the cascade is $\sim e^{-\tau_{max}}$,
then the mass will be
\begin{equation}  
\label{massis}
M_{\rm saxion} \sim {g_{s} \over \sqrt{\tau_{max}}} ~\Lambda ~. 
\end{equation}

Therefore, as long as one works in a compactification where this mass scale
is larger than the other effects communicating the SUSY-breaking auxiliary
field VEV to the saxion (which is not always true), 
the Higgs mechanism acts to help remove this possible source
of instability.
This remark is relevant to the Goldstone/saxion multiplets
arising from the breaking of each $U(1)_B$
that occurs in our theory.

However, this argument is far from conclusive.  There are also expected
to be contributions to the mass arising from the strongly coupled dynamics.
If for instance $\phi$ is the multiplet containing the saxion (as its real
part), and ${\cal M}$ is the multiplet that gets the SUSY-breaking F-term,
terms of the form
\begin{equation}
\label{trouble}
\int d^4\theta {c \over \Lambda^2} {\cal M}^{\dagger} {\cal M} (\phi +
\phi^{\dagger})^2
\end{equation}
should be expected to arise in the K\"ahler potential, with
$c$ some number which is a priori of ${\cal O}(1)$.
Obviously depending on the sign of $c$ this contribution can either help
stabilize or try to de-stabilize the saxion; for the ``wrong sign'' of $c$,
only moderately small $|c|$ can be overcome by the contribution
(\ref{massis}). Here, we expect $F_{\cal M} \sim
m\Lambda$, so for small quark masses, the contribution \eref{massis} can plausibly dominate.

Note that in the supersymmetry breaking states of \cite{Kachru:2002gs}, the saxion
indeed gets stabilized around zero. This is most convincingly shown
by computing the anti-D3 brane tension, which is minimum for vanishing 
saxion as shown in Figure 6 of \cite{Dymarsky:2005xt}. As we will review in \S\ref{gravity_section}, our
set up is slightly different because we are no longer in the probe approximation for the anti-D3-branes.

%=====================================================================+
\subsection{Meta-stable vacuum in the $\IZ_2$ orbifold of the conifold}
%=====================================================================+

We now come back to the theory at the bottom of the cascade for the
$\IZ_2$ orbifold of the conifold, Figure \ref{quiver_Z2_conifold_2} and superpotential
\eref{W_Z2_conifold_2}. We concentrate here on the case where $P=M$.

We consider that both nodes 1 and 3
have confining dynamics. Indeed, they should be the ones reaching strong
coupling first. We can actually assume that
$\Lambda_1, \Lambda_3\gg \Lambda_2$.
The relative hierarchy between
$\Lambda_1$ and $\Lambda_3$ is not fixed for the moment.
We thus have the following effective fields
\beq
{\cal M}=X_{21}X_{12}~, \quad {\cal N}=X_{23}X_{32}
\eeq
\beq
{\cal B} = (X_{12})^P, \quad \tilde {\cal B} =
(X_{21})^P, \quad {\cal C}=(X_{32})^P, \quad \tilde {\cal C}=(X_{23})^P~.
\eeq

The effective superpotential, which includes the tree level piece
\eref{W_Z2_conifold_2} (note that cyclic permutations among fundamental fields are allowed in the superpotential,  
since a trace on gauge indices is always understood) and the quantum constraints for nodes 1 and 3, read
\beq
W=h{\cal M} {\cal N}+\lambda_1(\det {\cal M}-{\cal B} \tilde {\cal B}-\Lambda_1^{2P})
+\lambda_3(\det {\cal N} - {\cal C} \tilde {\cal C}-\Lambda_3^{2P})~.
\eeq
 
Treating node 2 as classical, we can integrate out all the effective fields.
The F-terms that have to vanish in a supersymmetric vacuum are
\beq
\det {\cal M}-{\cal B} \tilde {\cal B}=\Lambda_1^{2P}~, \qquad \det {\cal N} -{\cal C} \tilde {\cal C}=\Lambda_3^{2P}
\label{F_1}
\eeq
\beq
h{\cal N} + \lambda_1(\det {\cal M}){\cal M}^{-1}=0~, \qquad h {\cal M} +\lambda_3(\det {\cal N}){\cal N}^{-1}=0
\eeq
\beq
\lambda_1 {\cal B}=0=\lambda_1\tilde {\cal B}~, \qquad \lambda_3 {\cal C}=0=\lambda_3 \tilde {\cal C}~.
\eeq

First of all, we can go on the baryonic branch for both nodes 1 and 3.
This implies having $\lambda_1=0=\lambda_3$, which in turn sets to zero the
VEVs of the two mesons ${\cal M}$ and ${\cal N}$. Hence we have two decoupled
one-dimensional baryonic branches, and eventually a confining $SU(P)$ SYM
at node 2. This should correspond on the gravity side to a single deformation.

A second choice is to be on the mesonic branch at both nodes 1 and 3.
Then we need $\lambda_1,\lambda_3\neq 0$ and hence we must set all baryonic
VEVs to zero. The mesons are both of maximal rank due to the quantum constraints,
and are eventually related by ${\cal N} \sim \Lambda_1^2 \Lambda_3^2 {\cal M}^{-1}$.
Hence we have only one mesonic moduli space. Moreover, node 2 is higgsed
to $U(1)^{P -1}$ and does not reach strong coupling.
The gravity dual interpretation of the above vacuum is that the ${\cal N}=2$
fractional branes are exploring their moduli space.

We could now consider having one node
on the baryonic branch and the other on the mesonic branch.
However, putting, say, node 1 on the mesonic branch would require ${\cal M} \neq 0$
while putting node 3 on the baryonic branch implies $\lambda_3=0$,
which is not consistent with the vanishing of
the F-term $\partial W/\partial{\cal N}=0$.
If all the other F-terms are vanishing, we would have a vacuum energy
\beq
V=|\Lambda_3|^2\sum_{i=1}^P|h m_i|^2
\label{V_vacuum_1}
\eeq
where $m_i$, $i=1,\ldots,P$, are the eigenvalues of ${\cal M}$ and
we have assumed that $K^{{\cal N}\bar {\cal N}}\sim |\Lambda_3|^2$.
The F-term on the left of \eref{F_1} constrains the eigenvalues of ${\cal M}$
according to $\det {\cal M}=\prod_{i=1}^P m_i=\Lambda_1^{2P}$. Minimizing \eref{V_vacuum_1}
subject to this constraint, we conclude that the $m_i$ are classically stabilized
at $m_i=\Lambda_1^2$ for all $i$. The vacuum energy at the meta-stable vacuum then becomes
\beq
V_{meta}=P |h\Lambda_1^2|^2 |\Lambda_3|^2~.
\eeq

We could do the reasoning in two steps. For example, if $\Lambda_1 \gg \Lambda_3$, we can first integrate out
the dynamics of node 1. If it is on the mesonic branch, the F-conditions
on the left column tell us that $h{\cal M} {\cal N} = P h \Lambda_1^2(\det {\cal N})^{1/P}$.
Hence we arrive at the superpotential
\beq
W=P h \Lambda_1^2(\det {\cal N})^\frac{1}{P} +
\lambda_3(\det {\cal N}- {\cal C}\tilde {\cal C}-\Lambda_3^{2P})~.
\label{w1}
\eeq

Integrating now over the dynamics at node 3, we
recover in particular the F-term for ${\cal N}$ that reads
\beq
h \Lambda_1^2 (\det {\cal N})^\frac{1}{P} {\cal N}^{-1} +\lambda_3(\det {\cal N}) {\cal N}^{-1}=0~.
\eeq

We do get supersymmetric vacua on its mesonic branch $\lambda_3\neq 0$,
while we get non-supersymmetric states on the
baryonic branch, where we assume that $(\det {\cal N})^\frac{1}{P} {\cal N}^{-1} \sim 1$.
Their meta-stability should be argued as in the SQCD case.
Note that what plays the r\^ole of the mass is $h\Lambda_1^2$,
hence the ISS regime of small mass compared to the dynamical scale should be attained for 
$h\Lambda_1^2 \ll \Lambda_3$.
We will describe why we think it is possible to tune the D-brane couplings to 
attain such a regime after we 
discuss more details of the Calabi-Yau geometry in the next section. 

In the non-supersymmetric states, the mesonic VEVs actually leave a
left-over $U(1)^{P-1}$ gauge symmetry which would be classically
enhanced to $SU(P)$ because all eigenvalues coincide at the minimum. 
Quantum effects should however prevent this from happening, along the lines of 
\cite{Johnson:1999qt}.

%%%%%%%%%%%%%%%%%%%%%%%%%%%%%%%%%%%%%%%%%%%%%%%%%%%%%%
\section{Gravity dual}
%%%%%%%%%%%%%%%%%%%%%%%%%%%%%%%%%%%%%%%%%%%%%%%%%%%%%%
\label{gravity_section}

In this section, we describe the gravity dual to the field theory
we have studied in the previous sections.
The $\IZ_2$ orbifold of the conifold is described by equation \eref{equation_Z2_conifold}, which we re-write below
for convenience
\begin{equation}
\label{geo}
(xy)^2 = zw~.
\end{equation}

As we have seen, the gauge theory (for $P=M$, which is the case we will focus on from now on)
has three different interesting classes
of vacua: i) the baryonic branch on nodes 1 and 3, which exhibits
confinement and chiral symmetry breaking, ii) the mesonic branch on nodes
1 and 3, which gives rise to a Coulomb phase with gauge group $U(1)^{P-1}$,
and iii) the mixed branch, where one finds meta-stable dynamical supersymmetry
breaking vacua.

For class i), we expect a gravity dual with a
geometric transition description along the
lines of \cite{Klebanov:2000hb} (see also \cite{Gopakumar:1998ki,Vafa:2000wi}), where the branes disappear
and are replaced by fluxes.
The IR physics is then captured by the flux superpotential \cite{Gukov:1999ya}
in a smooth background geometry.

The solutions in class ii) will instead have a description with explicit
probe D5 branes wrapped on a $\IP^1$.  The geometrical moduli space of
the $\IP^1$ (raised to an appropriate power and symmetrized),
reproduces the moduli space of the gauge theory.

Finally, for class iii), we will propose a gravity dual
which incorporates and generalizes the strategy of \cite{Kachru:2002gs}.

We will see that the flux superpotentials in the appropriate
deformations of (\ref{geo}) reproduce the low-energy
Taylor-Veneziano-Yankielowicz (TVY) superpotential \cite{Taylor:1982bp}
of the coupled super-QCD theories.

%======================================================================
\subsection{Effective superpotential and basic properties of the geometry}
%======================================================================

\label{section_VY}

On the gauge theory side, we can derive the low-energy superpotential
as follows.
Recall that for $SU(N_c)$ SQCD with $N_f$ flavors and meson superfield
${\cal M}$, the
effective superpotential is \cite{Taylor:1982bp}
\begin{equation}
\label{effsup}
W = - (N_c - N_f)  S + S {\rm log}\left( {S^{N_c-N_f} \over \Lambda^{3N_c - N_f}}
{\rm det} {\cal M}\right)~.
\end{equation}

There is actually an ambiguity in the linear term in the glueball superfield $S$, coming from
the possibility of shifting the bare gauge coupling; we will use this
freedom to fix the coefficient in a way that makes sense from the dual
gravity perspective.
In the case that the flavors become massive with (non-degenerate) mass
matrix $m$, by integrating
them out and matching, one easily sees that the new effective superpotential
should take the form
\begin{equation}
\label{effsup2}
W \to - N_c S + S{\rm log} \left( {S^{N_c} \over \Lambda^{3N_c - N_f} {\rm det} m}
\right)~.
\end{equation}

We note that in the case $N_c = N_f$, (\ref{effsup}) reproduces
the quantum deformed mesonic branch -- $S$ acts as a Lagrange multiplier
enforcing the condition
\begin{equation}
{\rm det} {\cal M} = \Lambda^{2N_c}~.
\end{equation}

The gauge theory at the end of our cascade
is more complicated than massive super-QCD, consisting of three
interacting gauge sectors.
However, it simplifies in various limits.
In the case that we go to the
baryonic branch of nodes 1 and 3, i.e.
class i) of the solutions above, the low-energy effective theory simply
consists of a pure $SU(P)$ gauge theory arising from node 2.
The effective superpotential we expect, by analogy with \cite{Klebanov:2000hb}, is
\begin{equation}
\label{classone}
W = \frac{k}{g_s} S_2 + P S_{2} {\rm log} S_2~.
\end{equation}
We fixed the coefficient of the term linear in $S_2$ 
in a way that will match the gravity
expectations, as we explain below.
Solving $\partial_{S}W = 0$ yields the $P$ vacua characteristic of
gaugino condensation in pure ${\cal N}=1$ $SU(P)$ gauge theory.

We can also write a simple model superpotential for the class ii) solutions.
Let us imagine working in the regime where $\Lambda_2$ is very small
(so the $SU(P)$ symmetry of node 2 is viewed as a global symmetry).
The gauge groups at nodes 1 and 3 both have $N_f = N_c$.
From the quartic superpotential,
we furthermore see that the effective mass matrix $m$ for the quarks at node 3
(which we take to be the node eventually responsible for DSB) 
is the $P \times P$ meson matrix of node 1, ${\cal M} = X_{21}X_{12}$.

Then we expect an effective superpotential describing 
the mesonic branch of node 1 to be
\begin{equation}
\label{nodetwo}
W_1 = S_{1} {\rm log} \left( {{\rm det} {\cal M} \over \Lambda_1^{2P}} \right)~.
\end{equation}
Similarly, the effective superpotential describing
the glueball superfields
associated to node 3 will be
\begin{equation}
\label{nodeone}
W_3 =\frac{k}{g_s}  S_3 + S_3 {\rm log} \left({S_3^P \over {\Lambda_{3}^{2P} h^P {\rm det}
{\cal M}}} \right) ~.
\end{equation}
The total effective superpotential is
\begin{equation}
\label{classtwo}
W_{tot} = W_1 + W_3
\end{equation}
and provides a coupling between the two sectors via the dual role of the
${\cal M}$ matrix, which is a meson superfield for node 1 and 
a flavor mass term for node 3. Note that the symmetry between nodes 1 and 3
is restored once we extremize the above superpotential, which amounts
to being in the supersymmetric states ii).

It is interesting to ask, how should one derive the superpotentials
(\ref{classone}) and (\ref{classtwo}) directly on the gravity side?
Namely, we should look for a set of fluxes and geometric moduli that reproduce
the above superpotentials via the
flux-induced Gukov-Vafa-Witten  (GVW) superpotential \cite{Gukov:1999ya}
\beq
\label{GVWsuper}
W \propto \int G_3 \wedge \Omega
\eeq
where $G_3 = F_3 - \tau H_3$ and for simplicity we fix the IIB axio-dilaton
to be $\tau=\frac{i}{g_s}$. 
We first focus on the more involved (\ref{classtwo}) and then make
some comments about the simpler (\ref{classone}).
Our singularity has three independent 2-cycles, as can be seen most easily
from the toric web diagram explained in Appendix \ref{appendix_toric}.
We will call ${\cal C}_1$ the cycle over which one 
wraps the fractional brane
corresponding to the rank assignment $(1,0,0,0)$ on the quiver.
Similarly, we call ${\cal C}_3$ the 2-cycle corresponding to the
$(0,0,1,0)$ brane. Lastly, a convenient choice is to call  ${\cal C}_2$ the cycle corresponding
to the rank assignment $(0,1,1,0)$.\footnote{The asymmetry between nodes 1 and 3 with 
respect to the cycle ${\cal C}_2$ might be disturbing 
for the reader. A more symmetric situation 
could be achieved by identifying the 2-cycles on the toric web after 
performing a flop transition on one of the two conifold singularities.
However, our choice of basis is the one making the following arguments the
clearest.} Each of these 2-cycles can be viewed as the base of B-cycles $B_1, B_2$ and $B_3$.
These B-cycles are noncompact, but we will imagine compactifying them as in
\cite{Giddings:2001yu}. Alternatively, we could work with a long distance cut-off $\rho_c$
on the noncompact geometry, which is mapped to a renormalization scale $\mu$ by the usual
relation $\rho_c=2\pi l_s^2\mu$, and would define for us the bare gauge theory parameters.
There is also a dual basis of three compact 3-cycles, the A-cycles
$A_1, A_2$ and $A_3$.

Since the branes wrapping on the 2-cycles ${\cal C}_1$ and
${\cal C}_3$ are deformation branes, we expect the dual 3-cycles
$A_1$ and $A_3$ to have moduli controlling their deformations.
These basically describe two conifold singularities.
The periods in such conifold geometries satisfy
\begin{equation}
\int_{A_i} \Omega = z_i~,~\int_{B_i} \Omega = {z_i\over 2\pi i} ~{\rm log}(z_i) +
{\rm regular} \qquad i=1,3~.
\label{compmod}
\end{equation}

We identify the $z_1$ and $z_3$ coordinates on the moduli space with the
glueball superfields $S_1$ and $S_3$ above. 
The brane wrapped on the 2-cycle ${\cal C}_2$ on the other hand is an 
${\cal N}=2$ fractional brane, and hence, much as in the $\IC\times\IC^2/\IZ_2$
geometry, its dual $A_2$ cycle does not deform, 
and the modulus
associated to it vanishes on-shell.

We expect that the superpotential (\ref{classtwo}) should be derived by
choosing appropriate RR and NS three-form fluxes. Due to the complete symmetry between node 1 
and node 3 in the $P=M$ case we are considering, several different fractional brane bases are 
equivalent for reproducing the superpotential we are after. In what follows, we choose for convenience the  
basis where the rank assignment in \fref{quiver_Z2_conifold_2} is obtained considering $P$ fractional branes of 
type $(0,0,1,0)$ and $P$ fractional branes of type $(1,1,0,0)$, the latter corresponding to the cycle 
${\cal C}_4= {\cal C}_1+{\cal C}_2-{\cal C}_3$ (looking at the toric web in Appendix B, \fref{toric_web}, 
one can easily recognize that such cycle indeed corresponds to an ${\cal N}=2$ fractional brane). This 
choice enables us to get the superpotential \eref{classtwo} more straightforwardly. We propose
\begin{equation}
\label{F3b}
\int_{A_3} F_3 = P
\end{equation}
and
\begin{equation}
\int_{B_3} H_3 = -k
\end{equation}
as the choices of RR and NS flux, respectively. This is the expected
geometric transition, arising from adding $P$ branes of the $(0,0,1,0)$ kind. 
The dependence on the meson field is reproduced if one adds $P$ probe D5 branes of ${\cal N}=2$ type
to the geometry; as anticipated, we want to add $P$ $(1,1,0,0)$ branes, whose corresponding cycle is 
${\cal C}_4={\cal C}_1+{\cal C}_2-{\cal C}_3$.

The meson then appears in the superpotential
in a way fixed by a standard disc computation, described in
\cite{Berenstein:2003fx,Imeroni:2003cw}, and precisely reproducing the field
theory result.
Let us explain in some more detail how the meson dependence arises in the flux superpotential. Our arguments
follow closely the ones presented in \cite{Imeroni:2003cw} for a similar model.
There it is shown that ${\cal N}=2$ fractional branes wrapped on a 2-cycle
$\cal C$ which is at the base of a (non-compact) B-cycle, when they 
are scattered on their moduli space, give rise to the following fluxes
\beq
\int_A G_3 = 0 , \qquad \int_B G_3 = i \log (\det {\cal M})
\eeq
where $\cal M$ parametrizes the positions
of the D5-branes along the curve of $A_1$ singularities.\footnote{We use 
$A_1$ to denote a compact 3-cycle as well as
a $\IC^2/Z_2$ singularity. We hope the reader can easily discern the meaning from the context.}

Thus, we expect the fluxes in our geometry to change accordingly
\beq
\begin{array}{ccc}
\int_{B_1} G_3 =  0 &  \ \ \rightarrow \ \ & \int_{B_1} G_3 = i\log (\det {\cal M}). \\ \\
\int_{B_3} G_3 = i\frac{k}{g_s} & \ \ \rightarrow \ \ &\int_{B_3} G_3 = 
i\frac{k}{g_s}\, -i\log (\det 
{\cal M}) 
\end{array}
\label{G3_probes_1}
\eeq
The integral of $G_3$ over $B_2$, though non vanishing, does not enter
the superpotential because the integral of $\Omega$ over $A_2$ is vanishing.
A word of caution has to be said regarding the integrals over the non-compact
B-cycles. Besides the UV cut-off, there is an 
additional short-distance cut-off that accounts for the break-down 
of the gravity
approximation very close to the D5-branes. We have not written this 
cut-off dependence explicitly on the right hand side 
of the equations (\ref{G3_probes_1}), but it can be shown
to combine with the data above 
to give the expected result in the superpotential. Indeed, plugging the supergravity fluxes 
\eref{F3b},\eref{G3_probes_1} and the geometric periods \eref{compmod} into \eref{GVWsuper} we get the expected 
field theory result \eref{classtwo}. 

Let us now briefly discuss the class i) solutions.
The superpotential \eref{classone} clearly indicates that there should
be $F_3$ flux through only one of the 3-cycles, namely the one dual
to the ${\cal C}_2-{\cal C}_3$ cycle, over which 
one wraps a $(0,1,0,0)$
fractional brane. There is also
$H_3$ flux through its corresponding non-compact cycle $B_2-B_3$. 
The other cycle $A_1$,
having no 3-flux, can shrink to zero size without introducing singularities
in the full supergravity solution (much as in the conformal conifold theory
\cite{Klebanov:1998hh}).
We then just have to identify the modulus $z$ controlling the size of the blown-up
3-cycle with $S_2$.\footnote{This modulus is essentially $z_3$ as
in \eref{compmod}. However in the present case it is associated
to $S_2$ rather than $S_3$. Indeed, it is the
initial charges and the particular vacuum one is choosing that selects
which scales $S_i$ are relevant and how they have to be identified with the
geometrical moduli.} As before, plugging the supergravity fluxes into the GVW 
formula we get the correct field theory superpotential \eref{classone}.

%====================================
\subsection*{Couplings and Scales}
%====================================

In \S2.1\ and \S4.2\ we discussed several regimes, such as 
$\Lambda_1 \gg \Lambda_3 \gg \Lambda_2$
and $\Lambda_3 \gg \Lambda_1 \gg \Lambda_2$. How can we obtain them?
Given the identification of cycles in the geometry and fractional branes discussed after 
\eref{GVWsuper}, it follows that the gauge couplings for the three gauge groups satisfy
\begin{equation}
{1\over g_{1,3}^2} \sim \int_{{\cal C}_{1,3}} B_2,~~
{1\over g_2^2} \sim \int_{{\cal C}_2 - {\cal C}_3} B_2~.
\end{equation}
It is clear that we have room for tuning the above quantities in order
to reach any such regime.

Recall however that we also want to have $h \Lambda_1^2 \ll \Lambda_3$.
We note that in the full 4-node quiver theory, there is an additional tunable
dimensionless parameter (which one can think of as $g_s$), 
but there are also two additional dimensionful couplings ($\Lambda_4$ and $h$).
As long as $h$ varies when one changes the additional parameter, 
this suffices to show that the various requirements we have placed on
the couplings and scales (for our analysis to apply) can indeed be met
in the string construction. 

%=====================================================
\subsection{Supersymmetric vacua}
%=====================================================

\label{section_SUSY_vacua}

We now describe how the supersymmetric vacua emerge from our
gravity description.  The minima of the flux superpotential for vacua
of class
ii) are easy to work out.  From the TVY superpotential
(\ref{classtwo}), integrating out the mesons,
one sees that the solutions lie where $S_1 = S_3$.
Hence, we expect the relevant geometry to describe two conifolds of equal
size.  The simple perturbation
\begin{equation}
\label{spacetwo}
(xy)^2 = zw \to (xy-\epsilon)^2 = zw
\end{equation}
accomplishes this, where $\epsilon$ is identified with the dynamical
scale of the confining gauge groups.
And from the identification of the GVW flux superpotential with the
TVY superpotential, we see that the deformation (\ref{spacetwo}) is
indeed the solution of the equations of motion for the complex
structure moduli -- we get two $S^3$ A-cycles which are deformed to
finite but equal size.

How do we incorporate the meson field?  The $U(1)^{P-1}$ gauge group
of the Coulomb branch of the $SU(P)$ node 2, is manifested in terms
of $P$ fractional probe D5 branes.  They wrap the small $\IP^1$ in
the curve of singularities visible in the geometry (\ref{spacetwo}),
located at $xy=\epsilon$.\footnote{This is precisely the cycle
${\cal C}_1+{\cal C}_2-{\cal C}_3$. Note that its dual A-cycle,
because of the on-shell relation $S_1=S_3$, 
has a vanishing integral of $\Omega$, as should
be the case for ${\cal N}=2$ fractional branes.}

The dual of the class i) vacua, is also
easy to describe.  The IR geometry is governed by the deformation
\begin{equation}
\label{spaceone}
(xy)^2 = zw \to (xy) (xy-\epsilon) = zw
\end{equation}
where $\epsilon$ is now related to the dynamical scale of the node 2
$SU(P)$ factor. There are no probe branes. 

The deformed geometries \eref{spacetwo}-\eref{spaceone} are
derived in appendix \ref{appendix_geometry} from the gauge theory.

%=====================================================
\subsection{Non-supersymmetric vacuum}
%=====================================================

How should we think about the meta-stable non-supersymmetric states of the
dual field theory?
In a somewhat similar context, involving the smooth gravity dual of
a confining gauge theory, it was observed in \cite{Kachru:2002gs} that one can
sometimes make meta-stable non-supersymmetric states by adding anti-brane
probes.  As long as the charges at infinity are fixed in the gravity
description, any such non-supersymmetric states must be interpreted as
particular vacuum
states in the supersymmetric gauge theory (at large 't Hooft
coupling).

In the case we have focused on, the three-node quiver with
occupation numbers $P-P-P$, the options are somewhat limited.
The gravity dual carries $N = kP$ units of D3 charge.  If we add an
anti-D3 probe, to maintain the same total charge, we would be forced to
add also a D3 probe, and the two would perturbatively annihilate.
In fact, the same situation holds if we add $2,3,\cdots,P-1$ anti-D3 probes.
However, the addition of $P$ anti-D3 probes introduces another option:
we can ``jump the fluxes'' so
\begin{equation}
\int_{A_3} F_3 = P~,~ \int_{B_3} H_3 = -k ~~\to~~
\int_{A_3} F_3 = P~,~\int_{B_3} H_3 = -(k+1)
\end{equation}
while adding the $P$ anti-D3 probes.  In this case, the total charges
at infinity are conserved. Therefore, this is another state in the same
supersymmetric theory we have been studying.

The mesonic branch characterized by the fluxes above also contains
$P$ D5 probes, wrapped around small cycles in the curve of $A_1$ singularities.
This is consistent with the left-over gauge symmetry present in the
supersymmetry breaking vacua of \S4.2.
By definition, at the quiver point in moduli space the fractional brane
charges are aligned with the D3 charges.  So the D5s will attract the
$P$ anti-D3 probes.  The result will be a state with the anti-D3 probes
dissolved in the D5s as gauge flux.
As described in \S4.2, the meta-stable states only exist in the case
$P=M$ in the field theory, and in that case, the preferred configuration
has $P$ equal eigenvalues of the meson matrix.  We therefore expect
the $P$ D5s to wrap a single small $\IP^1$ $\cal C$ in the locus of $A_1$
singularities, and manifest a worldvolume gauge field configuration with
\begin{equation}
\int_{\cal C} {\cal F} = -P
\end{equation}
which, via the Chern-Simons couplings in the D5 action, accounts for the
$P$ units of anti-brane charge.

This proposal matches nicely with the field theory.  In particular, it
is interesting that it is impossible to get meta-stable states by
adding $1,\cdots,P-1$ anti-branes, while adding $P$ leads to a natural
candidate.  This matches the fact that in the field theory, the only
(known) meta-stable supersymmetry breaking vacuum has an energy $\sim P$
in units of the dynamical scale.
It would be nice to find a precise gravity solution describing these
states.

Note that there are two equivalent meta-stable vacua, which result
from exchanging nodes 1 and 3. They are just mapped to each other by 
the $\IZ_2$ symmetry
of the geometrical background.

The reader may be confused about the distinction between the SUSY breaking
dynamics here, and that in \cite{Kachru:2002gs}.  There, the leading
effect on anti-D3 probes involved polarizing via the Myers effect 
\cite{Myers:1999ps} into
5-branes wrapping (contractible) $S^2$s.  
When there are few probe branes relative to the background RR flux,
the Myers potential exhibits meta-stable states.
Here, we claim there is a
stronger effect, which can yield bound states even for a number of
anti-D3s strictly comparable to the background flux.  
A heuristic argument in favor of this is as follows.
For large $P$, the three-form fluxes
are dilute, and the gradient of the Myers potential encouraging an anti-D3
to embiggen is very mild.  In this situation, it seems quite reasonable
that the attraction to the fractional D5s will provide a stronger
force on the anti-D3.  Indeed in flat space, this system is T-dual to
the D0-D2 system, which enjoys a long range attractive force and
exhibits a bound state which has binding energy that is an
${\cal O}(1)$ fraction of the original brane tension energies 
\cite{Polchinski:1998rr}.  If the probe anti-D3s are close enough to 
the D5s, this attractive force should be a stronger effect than the
force encouraging the anti-D3s to polarize. 
It is the presence of the background fractional D5s and their attraction
to the anti-D3s, that suggests to us that this system and the one in
\cite{Kachru:2002gs} behave differently. We consider this as supporting
evidence for the identification of the supersymmetry breaking states
on the gravity and gauge theory dual sides. 

We briefly note that in principle we have the possibility of adding a multiple
$nP$ of anti-D3 branes while shifting $\int_{B_3} H_3$ accordingly. We do not have a decisive argument against
stability (which is what we would expect, since these states are
not seen in the gauge theory), but we note that the binding energy
per unit anti-D3 probe decreases as the number of probes is increased,
so that eventually the Myers effect is likely to take over.

%=====================================================
\subsection{Comments on the full solution}
%=====================================================

After providing the correct fluxes reproducing the low energy effective dynamics of
the dual gauge theory, one might ask what the complete supergravity solution
describing our theory might be. This depends, of course, on which branches/vacua one is looking at.

The solutions characterizing branch i) in fact fit into a well known
general class of models. The self-dual 5-form satisfies a Bianchi identify
\begin{equation}
d\tilde F_5 = H_3 \wedge F_3 + \rho_{D3}
\end{equation}
where the first term on the RHS is the flux-induced D3 charge,
and the second term measures local charge density in probe D3 branes
(with an appropriate normalization).  For our setting, in the absence
of explicit probe branes, the complete D3 charge of $N$ is accounted
for by the fact that $\int H_3 \wedge F_3 = kP = N$.

The full IIB solution is very similar to the one discussed in
\cite{Klebanov:2000hb},
and falls in the general class of solutions described in \cite{Grana:2000jj,
Giddings:2001yu}.
The metric takes the form
\begin{equation}
\label{metricdef}
ds^2 = e^{2A(r)} \eta_{\mu\nu}dx^{\mu}dx^{\nu} + e^{-2A(r)}
\tilde g_{mn} dy^m dy^n
\end{equation}
with $\tilde g_{mn}$ the unit metric on the cone over an appropriate
Einstein manifold (in this case, the $\IZ_2$ orbifold of $T^{1,1}$), or its
deformation, in the non-conformal case; for class i) vacua this should be a metric on 
the deformed space \eref{spaceone}. The 5-form is determined in terms of the warp 
factor $A(r)$ via the equations
\begin{equation}
\tilde F_5 = (1+*) (d\alpha \wedge dx^0 \wedge dx^1 \wedge dx^2 \wedge dx^3)
\end{equation}
with
\begin{equation}
\alpha = e^{4A(r)}~.
\end{equation}

The three-form flux $G_3 = F_3 - \tau H_3$ is imaginary self-dual
\begin{equation}
*_6 G_3 = i G_3
\end{equation}
and purely of type (2,1).
$A(r)$ varies over a range dual to the range of scales covered in the
RG cascade, with
\begin{equation}
e^{A}\vert_{tip} = e^{-{4\pi k\over 3Pg_{s}}} ~.
\end{equation}

The axio-dilaton $\tau(y)$ is a constant in the background, which we can choose at infinity (in the compact
solutions, it is fixed by the flux superpotential and additional data involving fluxes on other cycles
in the compact manifold).

A fully backreacted supergravity solution for class ii) vacua is more difficult to achieve. We have again $P$ 
deformation branes which would provide a solution very similar to the one above, with the only difference that 
$\tilde g_{mn}$ should now be a metric on the deformed space \eref{spacetwo}. The problem is that there are $P$ 
${\cal N}=2$ branes around, too. They do not couple to the dilaton, which should then remain constant, and should 
also still provide an imaginary self-dual three-form flux. What is hard to find is the explicit form of the metric. 
Supergravity solutions for ${\cal N}=2$ fractional branes on undeformed orbifold-like singularities are well 
known \cite{Bertolini:2000dk}. However, in the present case one should compute their backreaction on the already  
deformed geometry \eref{metricdef}. It would be very interesting to find solutions of this kind, since 
they could play a r\^ole in several different contexts. However, such a challenge is beyond the scope of the present paper.

%%%%%%%%%%%%%%%%%%%%%%%%%%%%%%%%%%%%%%%%%%%%%%%%%%%%%%
\section{Conclusions}
%%%%%%%%%%%%%%%%%%%%%%%%%%%%%%%%%%%%%%%%%%%%%%%%%%%%%%

When leaving the realm of supersymmetric vacua, we are no longer guaranteed
that physics will not change qualitatively when parameters (such
as the 't Hooft coupling) are varied significantly. Hence a non-supersymmetric
meta-stable state which can be shown to exist when the parameter is small
(i.e. on the gauge side of the duality), might well not be visible anymore when the parameter
becomes large (i.e. on the gravity/string side of the duality), and vice-versa.

We have presented in this paper a simple example of a gauge/gravity dual pair
where we could provide evidence on both sides for the existence of
meta-stable states displaying dynamical supersymmetry breaking.
From the quiver gauge theory point of view, we are in a limiting case
of the theories studied by \cite{Intriligator:2006dd}. 
Meta-stability in the $N_f=N_c$ SQCD case that interests us
was briefly discussed in \cite{Intriligator:2006dd}, the most convincing argument
in favor of it being the relation, by decoupling of an additional
(more) massive flavor, to the more controlled case of $N_f=N_c+1$.
In the latter theory, meta-stability can be checked by a direct one-loop
computation. Following the meta-stable state as the mass of one
pair of chiral superfields is increased, we end up exactly on the ``baryonic
branch'' of the $N_f=N_c$ theory, the point that we have argued would be
the farthest from the SUSY vacua.
We consider this to be suggestive evidence that the states we have discussed
are not unstable in the gauge theory regime. 
Meta-stability could be set on a firmer footing by understanding the higher
order (in $1/\Lambda^2$) corrections to the K\"ahler potential.
In addition, identifying the optimal path from the meta-stable vacuum
to the SUSY vacua would give an estimate for
the lifetime of the meta-stable vacuum.

On the gravity side, we have provided geometrical arguments that
show that there are non-supersymmetric states
with the same quantum numbers as the field theory ones, and which
lack any obvious perturbative instabilities. 
We have a background geometry
\beq
(xy-\epsilon)^2=zw
\label{backgeo}
\eeq
which can be seen as being created by two sets of $P$ fractional branes.
It has a line of singularities which, in our set up,
supports another set of $P$ fractional branes,
of ${\cal N}=2$ kind. The latter can be expected to have a significant 
backreaction. Indeed, in the supersymmetric case, the system where all
the fractional branes do not explore their moduli space (but are on the
baryonic branch instead) should correspond
to the geometry where only one 3-cycle is blown up. On their mesonic branch,
the ${\cal N}=2$ fractional 
branes are still scattered as probes, though the geometry
deformed by the presence of sources is clearly more complicated.

The non-supersymmetric states correspond to the ${\cal N}=2$ fractional branes,
which are really D5-branes wrapped on small $\IP^1$s, carrying an additional
gauge flux with anti-D3 charge on the $\IP^1$.
The attraction between the anti-D3s and the wrapped D5-branes, and 
the precise matching of fluxes ($P$ anti-D3-branes
for $P$ D5-branes), makes stability plausible.

Note that in \cite{Kachru:2002gs}, a similar story was presented, but the
main difference there was that a probe computation revealed that, in order
to ensure meta-stability, the number
$p$ of anti-D3-branes must be much smaller than the number $P$ of fractional
branes originally creating the smooth geometry of the deformed conifold.
When $p$ is increased, the anti-D3-branes eventually polarize into
a big NS5-brane due to the Myers effect and decay perturbatively.
While in both cases for $P$ anti-D3-branes the probe approximation is clearly
not good, in the set up of this paper we could argue that there is a 
competing effect which can overcome the desire of the anti-D3s to
embiggen, namely their attraction towards the wrapped D5s. 
Hence, also on the gravity side, 
the non-supersymmetric states would naively be meta-stable.

Actually, we could imagine going further on the gravity side.
Not surprisingly, the geometry \eref{backgeo} is simply obtained from the
deformed conifold
\beq
xy-\epsilon=uv
\eeq
by performing the $\IZ_2$ orbifold acting as $u\rightarrow -u $,
$v\rightarrow -v $. Thus the full metric and fluxes of the background
geometry should be straightforwardly obtainable, through identifications
and the method of images, from the solution of Klebanov and Strassler
\cite{Klebanov:2000hb}. In order to describe
the meta-stable vacua we are after, we would then have to introduce
$P$ wrapped D5-branes (as, e.g., in \cite{Imeroni:2003cw}) with
the appropriate anti-D3 flux. This set up would not take into account
the (possibly large) backreaction of the additional probes, but should
already present the rough features of the system we want to describe.

For instance, we could be interested in the spectrum of gauge invariant
operators in this supersymmetry breaking vacuum. In particular,
we should find the massless fermion of broken supersymmetry, the goldstino.
Note that this massless mode should not be looked for among the
supergravity modes
(using, for instance, the methods of \cite{Argurio:2006my}),
but among the world-volume modes of the probe branes.
The situation is similar to \cite{Argurio:2006ew}, albeit for different
reasons.

Though we have focused on the simple example of the $\IZ_2$ orbifold
of the conifold throughout the paper, it seems likely that
one can find many similar cases.  Perhaps in some of them, one will
be able to find analogues of the more quantitatively accessible ISS vacua in the free magnetic range 
$N_c < N_f < {3\over 2} N_c$. A plausible 
way of achieving this goal is to consider our 3-node theory 
as a piece of a larger quiver as 
contemplated at the end of \S2.1. Such an extended theory could have additional 
gauge groups, more flavors for node 3 and appropriate superpotential interactions. The extra flavors may become 
massive dynamically by a mechanism similar to the one in \S2.1.

It would also be very interesting to find examples of meta-stable 
states similar to those analyzed in this paper in configurations where there is no 
stable supersymmetric vacuum, but instead (naively) a runaway behavior, 
such as
the $Y^{p,q}$ theories studied in \cite{Berenstein:2005xa,Franco:2005zu,Bertolini:2005di,Intriligator:2005aw,Brini:2006ej}. 

%%%%%%%%%%%%%%%%%%%%%%%%%%%%%%%%%%%%%%%%%%%%%%%%%%%%%%
\section*{Acknowledgements}
%%%%%%%%%%%%%%%%%%%%%%%%%%%%%%%%%%%%%%%%%%%%%%%%%%%%%%

We are happy to thank Sergio Benvenuti, Francesco Bigazzi, Gia Dvali,  Jarah Evslin, Gabriele Ferretti, Luciano 
Girardello, Yang-Hui He, Nissan Itzhaki, Igor Klebanov, Liam McAllister, Angel Uranga, Herman 
Verlinde and especially Ofer Aharony and Oliver DeWolfe for very helpful discussions.
R.A. and M.B. are partially supported by the European Commission FP6
Programme MRTN-CT-2004-005104, in which R.A. is associated to V.U. Brussel.
R.A. is a Research Associate of the Fonds National de la Recherche
Scientifique (Belgium). The research of R.A. is also supported by IISN - Belgium
(convention 4.4505.86) and by the ``Interuniversity Attraction Poles Programme --Belgian Science Policy''.
M.B. is supported by Italian MIUR under contract PRIN-2005023102 and by a
MIUR fellowship within the program ``Rientro dei Cervelli''.
S.F. is supported by the DOE under contract DE-FG02-91ER-40671. 
He would like to dedicate this work to Bel\'en Ariana Franco and her mom
Carla. The research of S.K. is supported by the DOE under contract DE-AC03-76SF00515,
the NSF under grant PHY-0244728, and a David and Lucile Packard Foundation
Fellowship for Science and Engineering.  He is grateful to the KITP for
kind hospitality while this research was being carried out.

\bigskip

%==============================================================================
%==============================================================================

\appendix

%%%%%%%%%%%%%%%%%%%%%%%%%%%%%%%%%%%%%%%%%%%%%%%%%%%%%%
\section{Geometry of the moduli space}
%%%%%%%%%%%%%%%%%%%%%%%%%%%%%%%%%%%%%%%%%%%%%%%%%%%%%%

\label{appendix_geometry}

The geometry of the moduli space can be derived from the chiral ring of the gauge theory with F-term relations,
both at the classical and quantum levels. In order to simplify the discussion we consider a single D3-brane probe
(in the full problem, the coordinates we discuss become the eigenvalues of mesonic operators and the moduli
space corresponds to the symmetric product of several copies of the same geometry).

The moduli space is determined by finding gauge invariant variables, writing down any algebraic relations
among them and then using F-term equations to simplify the relations even further.

Let us begin with the classical moduli space. We can define four quadratic gauge
invariant variables
\beq
\begin{array}{ccc}
x=X_{12}X_{21} & ~,~ & y=X_{23}X_{32} \\
z=X_{34}X_{43} & ~,~ & w=X_{41}X_{14}~.
\end{array}
\eeq

In addition, there are two quartic gauge invariants
\beq
u=X_{12} X_{23} X_{34} X_{41} ~~~,~~~ v=X_{14} X_{43} X_{32} X_{21}~.
\eeq

There is an algebraic relation among them
\beq
x y z w =u v~.
\label{algebraic}
\eeq

The F-term equations following from the tree-level superpotential \eref{W_Z2_conifold}
imply that $x=z$ and $y=w$. Hence, \eref{algebraic} becomes
\beq
(xy)^2=uv~.
\label{eq_orb_con}
\eeq

This is the classical geometry derived in \S\ref{section_orbifold_conifold} by the orbifold procedure.
We can now move on and derive the deformed geometries from the quantum gauge theories.

\begin{figure}[ht]
  \centering
  \includegraphics[width=4cm]{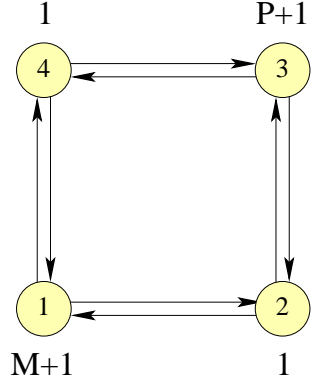}
  \caption{Quiver diagram for $M$ $(1,0,0,0)$ and $P$ $(0,0,1,0)$ branes after adding one D3-brane probe.}
  \label{quiver_probe_PM}
\end{figure}

A relevant case is when we have two sets of fractional branes,
say $M$ of type $(1,0,0,0)$ and $P$ of type $(0,0,1,0)$. 
In order to study the geometry, we add a single D3-brane probe. The resulting quiver is shown in \fref{quiver_probe_PM}.
The ``$SU(1)$'' nodes are not really gauge groups but have the effect of
adding chiral matter.\footnote{We adhere to the usual habit of neglecting
the $U(1)$ factors at every node as soon as non-trivial gauge dynamics
sets in.}

%%%%%%%%%%%%%%%%%%%%%%%%%%%%%%%
%%%%%%%%%%%%%%%%%%%%%%%%%%%%%%%

How is \eref{eq_orb_con} modified in the quantum theory? Nodes 1 and 3 have $N_f < N_c$ and hence ADS 
superpotentials are generated.
Starting again from \eref{algebraic}, we can study how it is deformed by 
using the F-terms that follow from $W_{tree}+W_{dyn}$.
The strong dynamics of nodes 1 and 3 is described in terms of their mesons (which we will later relate to 
the variables used to write
down the equation of the singularity) ${\cal M}_{ij}=X_{i1}X_{1j}$
and ${\cal N}_{ij}=X_{i3}X_{3j}$ with $i,j=2,4$.

Adding the tree-level and dynamical superpotentials, we obtain
\begin{eqnarray}
W&=& h\left({\cal M}_{22}{\cal N}_{22}+{\cal M}_{44}{\cal N}_{44}
-{\cal M}_{24} {\cal M}_{42}-{\cal N}_{24}{\cal N}_{42}\right) + \nonumber \\ & & +~
(M-1) \left( {\Lambda_1 ^{3M+1}\over \det {\cal M}}\right)^{1\over M-1}
+(P-1) \left( {\Lambda_3 ^{3P+1}\over \det {\cal N}}\right)^{1\over P-1}~.
\label{W3}
\end{eqnarray}

The F-terms altogether imply
\beq
\det {\cal M}=\left( {\Lambda_1 ^{3M+1}\over h^{M-1}}\right)^{1\over M}=\epsilon_1\quad, \quad
\det {\cal N}=\left( {\Lambda_3 ^{3P+1}\over h^{P-1}}\right)^{1\over P}=\epsilon_3 \label{detMN}
\eeq
and
\beq
{\cal M}_{22}={\cal N}_{44}\quad , \quad  {\cal M}_{44}={\cal N}_{22}~.
\label{F_terms_2}
\eeq
We have the following identifications
\beq
x={\cal M}_{22}\quad , \quad y={\cal N}_{22}\quad , \quad z={\cal N}_{44}\equiv x\quad , \quad w={\cal M}_{44}\equiv y
\eeq
where we have used \eref{F_terms_2}, and
\beq
u={\cal M}_{42}{\cal N}_{24}\quad ,\quad v={\cal M}_{24}{\cal N}_{42}~.
\eeq
We note that \eref{detMN} implies that ${\cal M}_{24}{\cal M}_{42}=
{\cal M}_{22}{\cal M}_{44}-\epsilon_1\equiv xy-\epsilon_1$,
and  ${\cal N}_{24}{\cal N}_{42}=
{\cal N}_{22}{\cal N}_{44}-\epsilon_3\equiv xy-\epsilon_3$~.
Using again \eref{F_terms_2},
we conclude that the deformed geometry corresponds to
\beq
(x y-\epsilon_1)(x y-\epsilon_3)=u v~.
\label{deformed_bb2}
\eeq
For generic deformation parameters $\epsilon_1$ and $\epsilon_3$, the above
geometry is completely regular. However, for $\epsilon_1=\epsilon_3$
it has a line of $A_1$ singularities on the locus
defined by $x y-\epsilon_1=0$.
On the locus of singular points, which contain small $\IP^1$s,
we can then add the fractional branes of ${\cal N}=2$ kind. This corresponds to the class ii) vacua 
discussed in \S\ref{gravity_section}. Note that the simplest way to satisfy the relation $\epsilon_1=\epsilon_3$
is when the ranks and scales are equal, $P=M$ and $\Lambda_1=\Lambda_3$. 

We can use these results also to determine the deformation of the geometry once we are in the baryonic branch for both nodes
1 and 3, which corresponds to the class i) vacua. This leaves a single $SU(M)$ confining gauge group, node 2, behind. 
As argued in \S\ref{section_VY}, this is expected to correspond to a single deformation. In fact, after rotating the 
quiver such that nodes are relabeled according to $(1,2,3,4)\rightarrow (4,1,2,3)$, the deformation follows from 
setting $P=0$ in our previous analysis. Hence, \eref{deformed_bb2} reduces to
\beq
(x y-\epsilon)(x y)=u v~.
\label{deformed_bb}
\eeq

It is straightforward to check that there is a remaining conifold singularity
in the moduli space. From a field theory point of view, this is expected to be
manifest when the dynamical scale of the leftover node goes to infinity (equivalently when $\epsilon \rightarrow \infty$).
This corresponds to studying the theory at energy scales much lower than $\Lambda_2$. Using rescaled variables
$\tilde{x}=\epsilon^{1/2} x$ and $\tilde{y}=\epsilon^{1/2} y$, \eref{deformed_bb} becomes the conifold equation $\tilde{x}\tilde{y}=u v$.

%%%%%%%%%%%%%%%%%%%%%%%%%%%%%%%%%%%%%%%%%%%%%%%%%%%%%%%%%%%%%%%%%
\section{Toric geometry, $(p,q)$ webs and complex deformations}
%%%%%%%%%%%%%%%%%%%%%%%%%%%%%%%%%%%%%%%%%%%%%%%%%%%%%%%%%%%%%%%%

\label{appendix_toric}

This appendix presents basic notions about toric singularities and their description
in terms of $(p,q)$ webs.
The goal is to provide a brief description of practical rules necessary to get an intuitive understanding
of the basic geometric features of a singularity and its possible deformations. We refer the reader to
\cite{Fulton,Aspinwall:1993nu,Morrison:1994fr} for more detailed presentations of toric geometry and to 
\cite{Leung:1997tw,Franco:2002ae} for comprehensive expositions
of the connections between web diagrams and toric geometry. These ideas are not crucial for understanding the 
current paper but are useful for identifying the cycles over which branes are wrapped. They also simplify the 
study of generalizations of our model.

A d-complex dimensional toric variety $V^d$ is
a generalization of a complex projective space, defined as
\beq
V^d=(\IC^n-F)/\IC^{*(n-d)}
\eeq
in which we quotient by $(n-d)$ $\IC^*$ actions and we remove a set of points $F$ in
order for such quotient to be well defined. The action of $\IC^{*(n-d)}$ on the
$\IC^n$ coordinates is defined by $(n-d)$ charge vectors $Q^a$ as
\beq
\lambda_a: (x_1,\ldots,x_n) \rightarrow (\lambda_a^{Q^a_1} x_1,\ldots,\lambda_a^{Q^a_n} x_n), \ \ \ \ \
a=1,\ldots, n-d~.
\eeq

The charges can be arranged into a matrix $Q=(Q^a_i)$. This approach to toric varieties
is known as the holomorphic quotient.

Alternatively, we can perform the quotient by $\IC^{*(n-d)}$ in two steps, decomposing each
$\IC^*=\IR^+ \times U(1)$. This approach is called the symplectic quotient.
We first fix the $\IR^{+(n-d)}$ levels via a moment map
\beq
\sum_{i=1}^n Q_a^i |x_i|^2=\xi_a, \ \ \ \ \ a=1,\ldots, n-d
\eeq
for some real parameters $\xi_a$. We refer to these equations D-terms. This is because they are actually
the D-terms of an $\mathcal{N}=2$ gauged linear sigma model (GLSM) with target space $\IC^n$ which reduces in
the infrared to a non-linear sigma model whose target space is the toric variety $V^d$ \cite{Witten:1993yc}.

Finally, we quotient by the $U(1)^{(n-d)}$ action defined by the charge matrix $Q$ (which gives the gauge groups
and corresponding charges of the GLSM).
Generic non-zero values of the $\xi_a$'s lead to a full resolution of the singularity. Setting
them to non-generic values (i.e. with some linear combinations equal to zero) produces a partial
resolution.

A simple way to represent a toric singularity is by means of a toric diagram. A toric diagram
for a d-complex dimensional toric variety is a set of points in the integer lattice ${\bf N}=\IZ^{(d)}$.
The toric diagram consists of $n$ vectors $\vec{v}_i$, $i=1,\ldots,n$. Each $\vec{v}_i$
represents an homogeneous coordinate $z_i$. The $\vec{v}_i$'s satisfy linear relations of the form
\beq
\sum_{i=1}^n Q_a^i \vec{v}_i=0
\eeq
with $Q_a^i \in \IZ$. In other words, $Q$ is given by the kernel of a matrix
whose columns are the $\vec{v}_i$'s. The matrix $Q$ computed this way, is precisely the
one that determines the $U(1)^{(n-d)}$ action of the symplectic quotient.

When the toric variety is Calabi-Yau, the toric diagram is simplified. In fact the manifold
is Calabi-Yau if and only if there exist a vector $\vec{h}$ in the lattice ${\bf M}$ dual to ${\bf N}$
such that
\beq
\langle\vec{h},\vec{v}_i\rangle=1 \ \ \ \ \forall \ \vec{v}_i~.
\label{condition_CY_toric}
\eeq

In other words, the toric diagram lives on an $(n-1)$-dimensional hyperplane at unit distance
from the origin. This means that the 3-complex dimensional toric Calabi-Yaus that we focus on
are represented by toric diagrams that are effectively 2-dimensional.

A related approach to toric varieties is to
construct them as n-dimensional torus fibrations $T^n$ over some base
spaces. For 3-complex dimensional toric varieties, the information about degenerations of the fibrations can 
be encoded in toric skeletons, also known as $(p,q)$ webs \cite{Aharony:1997ju,Aharony:1997bh}.

\begin{figure}[ht]
  \centering
  \includegraphics[width=9cm]{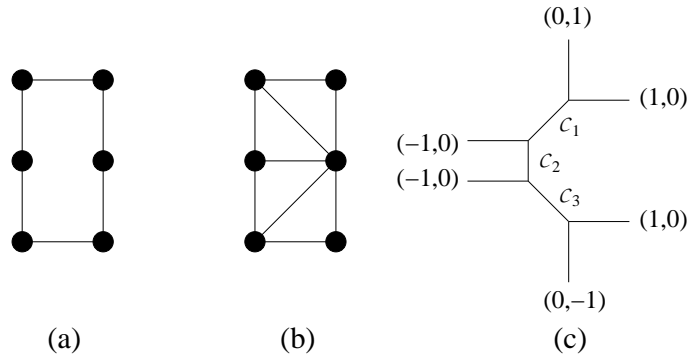}
  \caption{a) Toric diagram for the $\IZ_2$ orbifold of the conifold. b) Triangulation of the toric diagram. c) $(p,q)$ 
web with charges of external legs.}
  \label{toric_web}
\end{figure}

A web diagram is obtained by dualizing any triangulation
of the toric diagram. Edges, nodes and faces of the toric diagram are respectively mapped to
transverse edges, faces and nodes of the dual web.
Different triangulations of the toric diagram, and hence different forms of the dual web,
are related by flop transitions. \fref{toric_web} illustrates these ideas with the example considered along the paper.

%%%%%%%%%%%%%%%%%%%%%%%%%%%%%%%
%%%%%%%%%%%%%%%%%%%%%%%%%%%%%%%

Every segment in the web is characterized by a pair of integers, the so called  $(p,q)$ charges. These charges
determine the orientation of the segment $x:y=p:q$. In addtion, $(p,q)$ charge is conserved at
every node in the web.

There is a $T^2$ fibration over every point in the web diagram. The
additional circle action comes from the rotation on the phase of the
normal line bundle. Lines in the web diagram indicate where one of the
circles of the $T^2$ shrinks to zero. The entire $T^2$ vanishes at the
points where two lines meet. Then, finite segments in the web
correspond to $\IP^1$'s and compact faces represent compact 4-cycles.

In Figure \ref{toric_web}c), we have shown the basis of 2-cycles
that we refer to in \S\ref{gravity_section}. Note that fractional branes
wrapped on ${\cal C}_1$
and ${\cal C}_2$ have an orientation which can be represented by
circling the segment anti-clockwise, while a brane wrapped on ${\cal C}_3$ 
should correspond to the segment being circled clockwise.

\begin{figure}[ht]
  \centering
  \includegraphics[width=8cm]{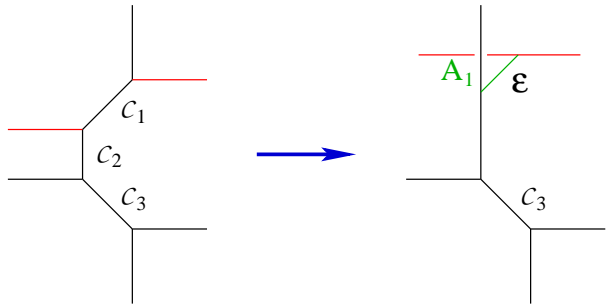}
  \caption{Deformation of the $\IZ_2$ orbifold of the conifold given by $(xy)(xy-\epsilon)=uv$. The green 
segment represents an $S^3$ with volume proportional to its length.}
  \label{toric_deformation_1}
\end{figure}

Web diagrams are very useful for identifying possible complex
deformations, in which an $S^2$ makes a transition into an $S^3$. They
correspond to decompositions of the web into sub-webs in equilibrium
(i.e. the sum of the $(p,q)$ charges of external legs vanishes for
each piece). \fref{toric_deformation_1} shows the web description of
the $(xy)(xy-\epsilon)=uv$ deformation, in which a single $S^3$ of
size $\epsilon$ is generated. The leftover diagram corresponds to a
conifold singularity, in agreement with the discussion in appendix
\ref{appendix_geometry}.

%%%%%%%%%%%%%%%%%%%%%%%%%%%%%%%
%%%%%%%%%%%%%%%%%%%%%%%%%%%%%%%

\begin{figure}[ht]
  \centering
  \includegraphics[width=9cm]{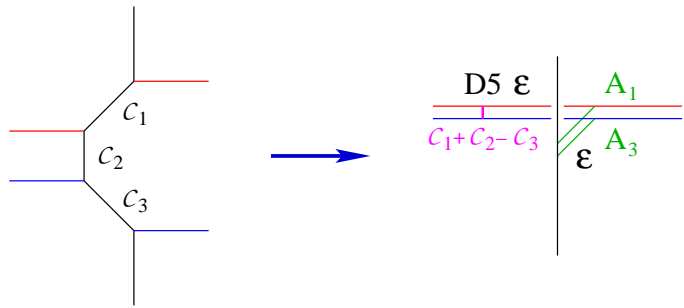}
  \caption{Deformation of the $\IZ_2$ orbifold of the conifold given by $(xy-\epsilon)^2=uv$. We show D5-branes 
wrapped over the $\IP^1$ along the curve of singularities in magenta.}
  \label{toric_deformation_2}
\end{figure}

As a final example, \fref{toric_deformation_2} provides a pictorial representation for the $(xy-\epsilon)^2=uv$ 
deformation. In this case, there are two $S^3$'s of equal size $\epsilon$. The Coulomb branch of node 2 of the gauge 
theory corresponds to motion of the D5-branes (indicated in magenta) wrapped over the $\IP^1$ along a curve of singularites, corresponding to
the cycle ${\cal C}_1+{\cal C}_2-{\cal C}_3$.

%%%%%%%%%%%%%%%%%%%%%%%%%%%%%%%
%%%%%%%%%%%%%%%%%%%%%%%%%%%%%%%

%%%%%%%%%%%%%%%%%%%%%%%%%%%%%%%%%%%%%%%%%%%%%%%%%%%%%%
\section{Type IIA description}
%%%%%%%%%%%%%%%%%%%%%%%%%%%%%%%%%%%%%%%%%%%%%%%%%%%%%%

\label{appendix_IIA}

Many features of gauge theories become geometrical when they are engineered using brane setups.
Recently, Type IIA configurations dual to meta-stable non-supersymmetric vacua of various gauge
theories were constructed \cite{Ooguri:2006bg,Franco:2006ht,Ahn:2006gn}. These construction allow a simple 
visualization of various
aspects of the field theories such as vacuum energy, pseudo-moduli 
and their stabilization by a  
1-loop effective potential. With this motivation, we proceed to construct the IIA dual of our
gauge theory.

\begin{figure}[ht]
  \centering
  \includegraphics[width=11cm]{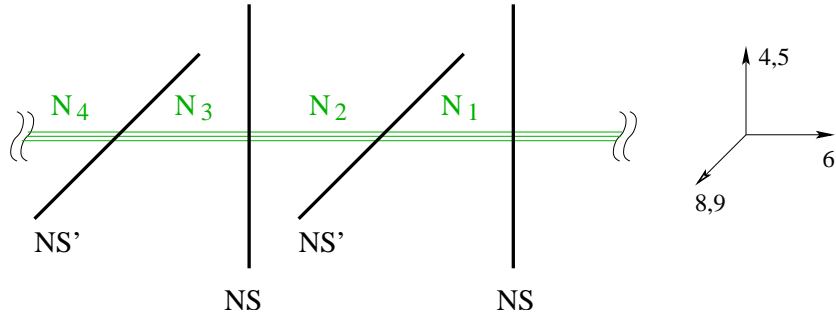}
  \caption{Type IIA brane configuration dual to the $(N_1,N_2,N_3,N_4)$ quiver. D4-branes are shown in green. The $x_6$ direction
is periodically identified.}
  \label{IIA_0}
\end{figure}

Regular and fractional D3-branes probing the $\IZ_2$ orbifold of the conifold are T-dual to a system of D4-branes
and relatively rotated NS5-branes in Type IIA. This configuration is shown, for arbitrary ranks of the gauge groups,
in \fref{IIA_0}. NS-branes extend in $(0,1,2,3,4,5)$, NS'-branes in $(0,1,2,3,8,9)$ and D4-branes in $(0,1,2,3,6)$. The
$x_6$ direction is compactified on a circle. Regular D3-branes become D4-branes extended along the entire $x_6$ circle,
while fractional D3-branes map to D4-branes stretched between NS5-branes.

%%%%%%%%%%%%%%%%%%%%%%%%%%%%%%%
%%%%%%%%%%%%%%%%%%%%%%%%%%%%%%%

\begin{figure}[ht]
  \centering
  \includegraphics[width=11cm]{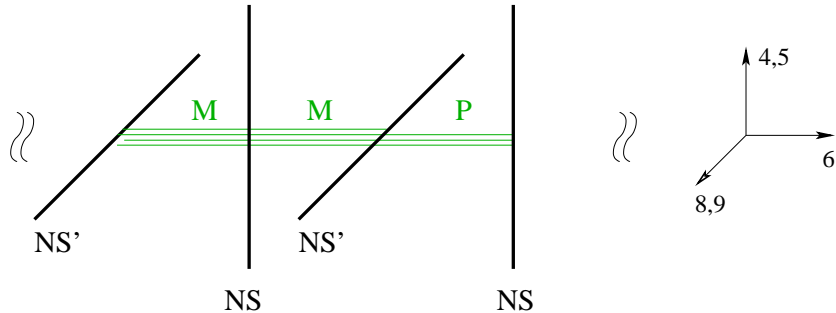}
  \caption{Type IIA brane configuration dual to the $(P,M,M,0)$ quiver.}
  \label{IIA_1}
\end{figure}

\fref{IIA_1} shows the configuration dual to the three node quiver of \fref{quiver_Z2_conifold_2}.

%%%%%%%%%%%%%%%%%%%%%%%%%%%%%%%
%%%%%%%%%%%%%%%%%%%%%%%%%%%%%%%

\begin{figure}[ht]
  \centering
  \includegraphics[width=11cm]{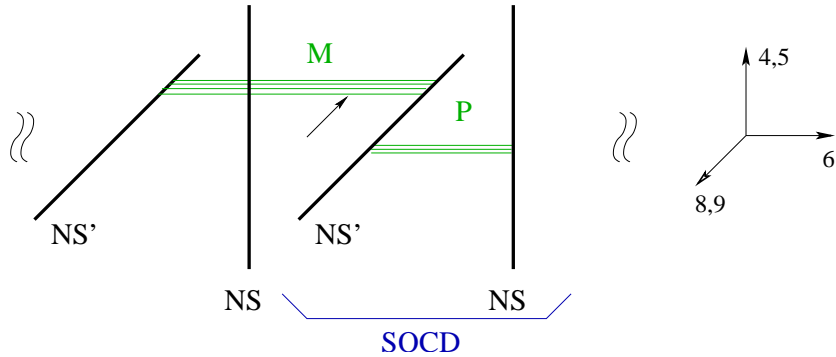}
  \caption{Type IIA configuration dual in the mesonic branch.}
  \label{IIA_2}
\end{figure}

The mesonic branch has a simple realization in the brane configuration.
It corresponds to combining the two sets of D4-branes at both sides of the left NS-brane and
moving them in the 89 directions as shown in \fref{IIA_2}. \footnote{What is not visible
is the dynamics that leads to a non-zero meson expectation value.} 
Interestingly, if we focus on the rightmost piece of the setup,
we identify the usual configuration dual to $SU(P)$ SQCD with massive flavors. \footnote{In fact
it is a trivial rotation of the way in which the configuration is usually presented in the literature.} 
The complex mass parameters correspond to the positions of the M D4-branes in 89. In this case, the mass matrix $m$ is
constrained by $\det m=h^P \Lambda_1^{2P}$ but the masses are dynamical.

%%%%%%%%%%%%%%%%%%%%%%%%%%%%%%%
%%%%%%%%%%%%%%%%%%%%%%%%%%%%%%%

\begin{figure}[ht]
  \centering
  \includegraphics[width=11cm]{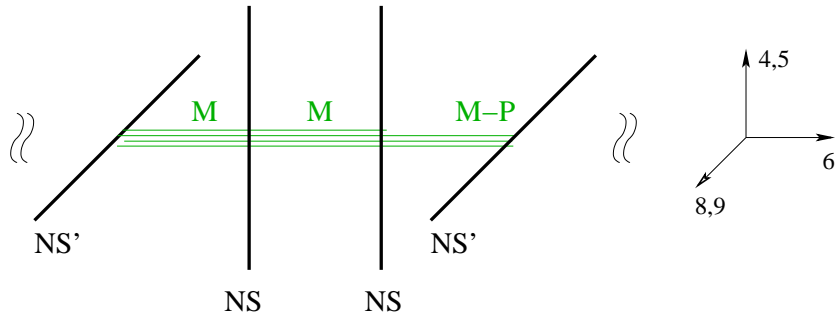}
  \caption{Type IIA configuration after Seiberg dualizing the $SU(P)$ node.}
  \label{IIA_3}
\end{figure}

The meta-stable vacuum is identified in the free magnetic theory obtained by performing
a Seiberg duality transformation on the $SU(P)$ gauge group. Let us perform that duality,
setting the meson vevs to zero for the time being. Seiberg duality corresponds to a continuation
through infinite coupling that is mapped in the IIA language to moving the NS' across the NS. The
resulting configuration is shown in \fref{IIA_3}.

%%%%%%%%%%%%%%%%%%%%%%%%%%%%%%%
%%%%%%%%%%%%%%%%%%%%%%%%%%%%%%%

\begin{figure}[ht]
  \centering
  \includegraphics[width=11cm]{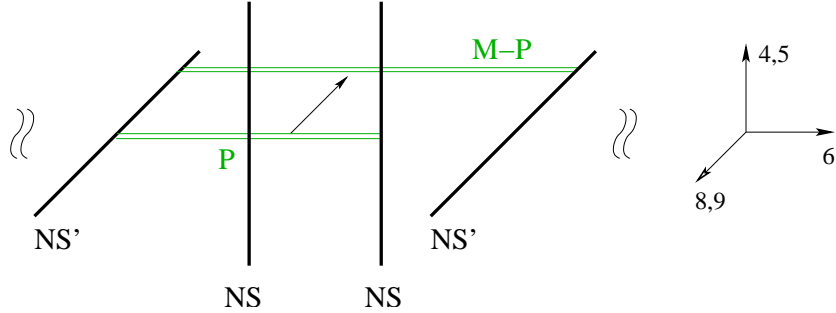}
  \caption{Type IIA configuration when the rank of the meson vevs is equal to $M-P$. Turning on additional non-zero
vevs break supersymmetry.}
  \label{IIA_4}
\end{figure}

Let us now reinsert the meson vevs. \fref{IIA_4} shows what happens when $M-P$ meson eigenvalues are non-zero.
In a T-dual version of the discussion in \S\ref{section_SUSY_vacua}, the system consisting of the two NS'-branes
and the $M-P$ D4-branes preserves ${\cal N}=2$ SUSY. Giving non-zero vevs
to the lowest $P$ components corresponds to moving the $P$ remaining D4-branes in the 89 directions. It is clear
that doing this breaks supersymmetry. 

%%%%%%%%%%%%%%%%%%%%%%%%%%%%%%%
%%%%%%%%%%%%%%%%%%%%%%%%%%%%%%%

Trying to extend these brane configurations to generic points in the MQCD parameter space will probably 
face the same obstacles studied in \cite{Bena:2006rg}. These subtleties are not present in our gauge/gravity setup.

%%%%%%%%%%%%%%%%%%%%%%%%%%%%%%%%%%%%%%%%%%%%%%%%%%%%%%%%%%%%%%%%

\bibliographystyle{JHEP}

\end{document}

%% file: texpref.tex
\newcommand{\eref}[1]{(\ref{#1})}

\newcommand{\fref}[1]{Figure~\ref{#1}}
\newcommand{\cref}[1]{Chapter~\ref{#1}}
\newcommand{\beq}{\begin{equation}}
\newcommand{\eeq}{\end{equation}}
\newcommand{\ba}{\begin{array}}
\newcommand{\ea}{\end{array}}
\newcommand{\bcenter}{\begin{center}}
\newcommand{\ecenter}{\end{center}}

\def\IB{\relax\hbox{$\inbar\kern-.3em{\rm B}$}}
\def\IC{\relax\hbox{$\inbar\kern-.3em{\rm C}$}}
\def\ID{\relax\hbox{$\inbar\kern-.3em{\rm D}$}}
\def\IE{\relax\hbox{$\inbar\kern-.3em{\rm E}$}}
\def\IF{\relax\hbox{$\inbar\kern-.3em{\rm F}$}}
\def\IG{\relax\hbox{$\inbar\kern-.3em{\rm G}$}}
\def\IGa{\relax\hbox{${\rm I}\kern-.18em\Gamma$}}
\def\IH{\relax{\rm I\kern-.18em H}}
\def\IK{\relax{\rm I\kern-.18em K}}
\def\IL{\relax{\rm I\kern-.18em L}}
\def\IP{\relax{\rm I\kern-.18em P}}
\def\IR{\relax{\rm I\kern-.18em R}}
\def\IZ{\relax\ifmmode\mathchoice
{\hbox{\cmss Z\kern-.4em Z}}{\hbox{\cmss Z\kern-.4em Z}}
{\lower.9pt\hbox{\cmsss Z\kern-.4em Z}}
{\lower1.2pt\hbox{\cmsss Z\kern-.4em Z}}\else{\cmss Z\kern-.4em Z}\fi}
\def\II{\relax{\rm I\kern-.18em I}}

\def\sCC{{\kern 0.27em\vrule height1.45ex width0.03em depth0em
          \kern-0.30em\rm C}}
\def\C{{\mathchoice
  {\sCC}
  {\sCC}
  {\kern 0.225em \vrule height1.05ex width0.025em depth0em \kern-0.25em \rm C}
  {\kern 0.180em \vrule height0.78ex width0.02em depth0em \kern-0.2em \rm C}
        }}
\def\sHH{{\rm I\kern-.16em{}H}}
\def\H{{\mathchoice
  {\sHH}
  {\sHH}
  {\rm I\kern-.13em{}H}
  {\rm I\kern-.13em{}H} }}
\def\sNN{{\rm I\kern-.16em{}N}}
\def\N{{\mathchoice
  {\sNN}
  {\sNN}
  {\rm I\kern-.12em{}N}
  {\rm I\kern-.10em{}N} }}
\def\sPP{{\rm I\kern-.16em{}P}}
\def\P{{\mathchoice
  {\sPP}
  {\sPP}
  {\rm I\kern-.12em{}P}
  {\rm I\kern-.10em{}P} }}
\def\sQQ{{\kern 0.27em \vrule height1.45ex width0.03em depth0em
          \kern-0.30em \rm Q}}
\def\Q{{\mathchoice
        {\sQQ}
        {\sQQ}
  {\kern 0.225em \vrule height1.05ex width0.025em depth0em \kern-0.25em \rm Q}
  {\kern 0.180em \vrule height0.78ex width0.020em depth0em \kern-0.20em \rm Q}
        }}
\def\sRR{{\rm I\kern-0.16em{}R}}
\def\R{{\mathchoice
  {\sRR}
  {\sRR}
  {\rm I\kern-0.12em{}R}
  {\rm I\kern-0.10em{}R} }}
\def\sZZ{{\rm Z\kern-0.32em{}Z}}
\def\Z{{\mathchoice
  {\sZZ}
  {\sZZ} 
  {\rm Z\kern-0.3em{}Z}     %.3
  {\rm Z\kern-0.25em{}Z} }}  %.25
\def\ZZZ{{\rm Z\kern-0.24em{}Z}}
\def\sII{{\rm I\kern-0.16em{}I}}
\def\I{{\mathchoice
  {\sII}
  {\sII}
  {\rm I\kern-0.12em{}I}
  {\rm I\kern-0.10em{}I} }}

\def\inbar{\,\vrule height1.5ex width.4pt depth0pt}
\font\cmss=cmss10 \font\cmsss=cmss10 at 7pt

\def\smiley{\hbox{\large$\bigcirc$\hspace{-0.80em}\raise.2ex
\hbox{$\cdot\cdot$}\kern-.61em\lower.2ex\hbox{\scriptsize$\smile$}}\ }
\def\frowny{\hbox{\large$\bigcirc$\hspace{-0.80em}\raise.2ex
\hbox{$\cdot\cdot$}\kern-.635em\lower.2ex\hbox{\scriptsize$\frown$}}\ }

\def\I{{\rlap{1} \hskip 1.6pt \hbox{1}}}

\makeatletter
\let\hangafter\@hangfrom
\makeatother